\documentclass[a4paper,11pt]{article}
\pdfoutput=1 

\usepackage{jheppub,bm,mathtools,slashed} 
\usepackage[usenames,dvipsnames,svgnames,table,x11names]{xcolor}
\usepackage{extarrows}
\usepackage{easybmat}

\usepackage{nicematrix}

\usepackage[T1]{fontenc} 
\usepackage[utf8]{inputenc}

\renewcommand*{\backref}[1]{}
\renewcommand*{\backrefalt}[4]{%
  \ifcase #1 %
    \relax 
  \or
    {\scriptsize (page~#2).}%
  \else
    {\scriptsize (pages~#2).}%
  \fi%
}

\definecolor{light_blue}{rgb}{0.15, 0.35, 0.95}
\definecolor{kit_green}{rgb}{0, 
0.58823 
, 0.50980 
}

\usepackage{tikz}
\usepackage{pgfplots}
\pgfplotsset{compat=1.14}
\usepackage{tikz-3dplot}
\usetikzlibrary{intersections, calc,positioning,decorations.pathreplacing,decorations.pathmorphing,arrows,arrows.meta,patterns,pgfplots.fillbetween,math,matrix}

\usepackage{multirow}
\usepackage{booktabs} 
\usepackage{siunitx} 
\usepackage{tabularx} 
\newcommand{\bhline}[1]{\noalign{\hrule height #1}}

\newcommand{\dd}{\mathrm{d}}
\renewcommand{\i}{\mathrm{i}}

\def\nn{q_\mathrm{e}}
\def\mm{q_\mathrm{m}}

\newcommand{\eg}{\emph{e.g.}}
\newcommand{\ie}{\emph{i.e.}}

\newcommand{\yitp}{{
    \includegraphics[height=0.8em]{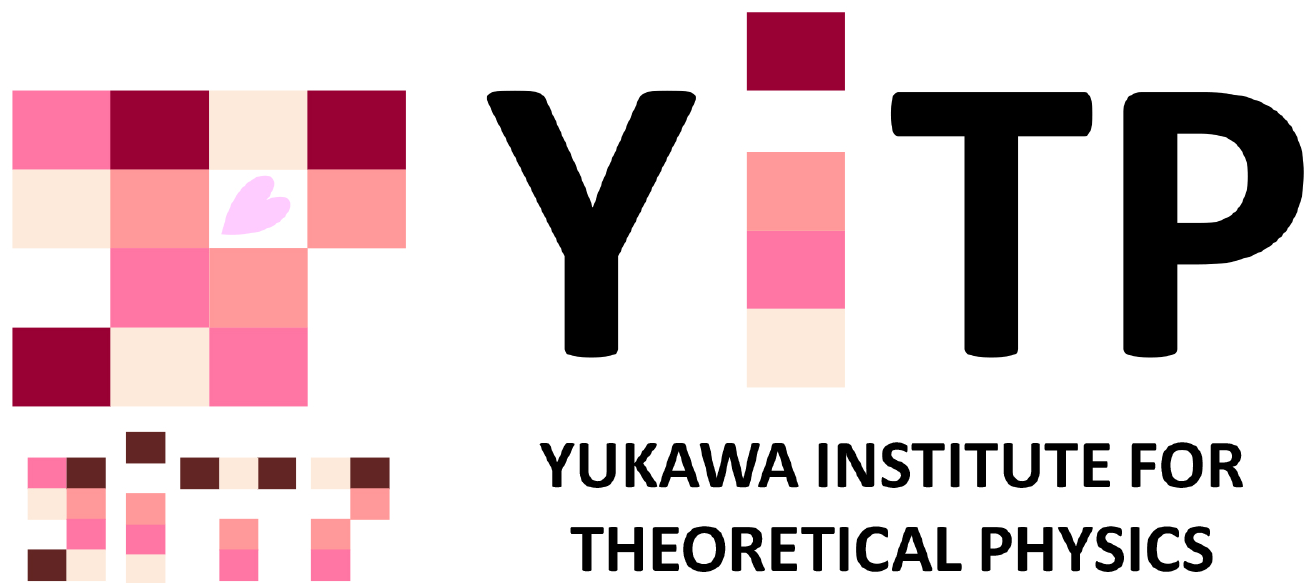}
}}
\newcommand{\kek}{{
    \includegraphics[height=0.8em]{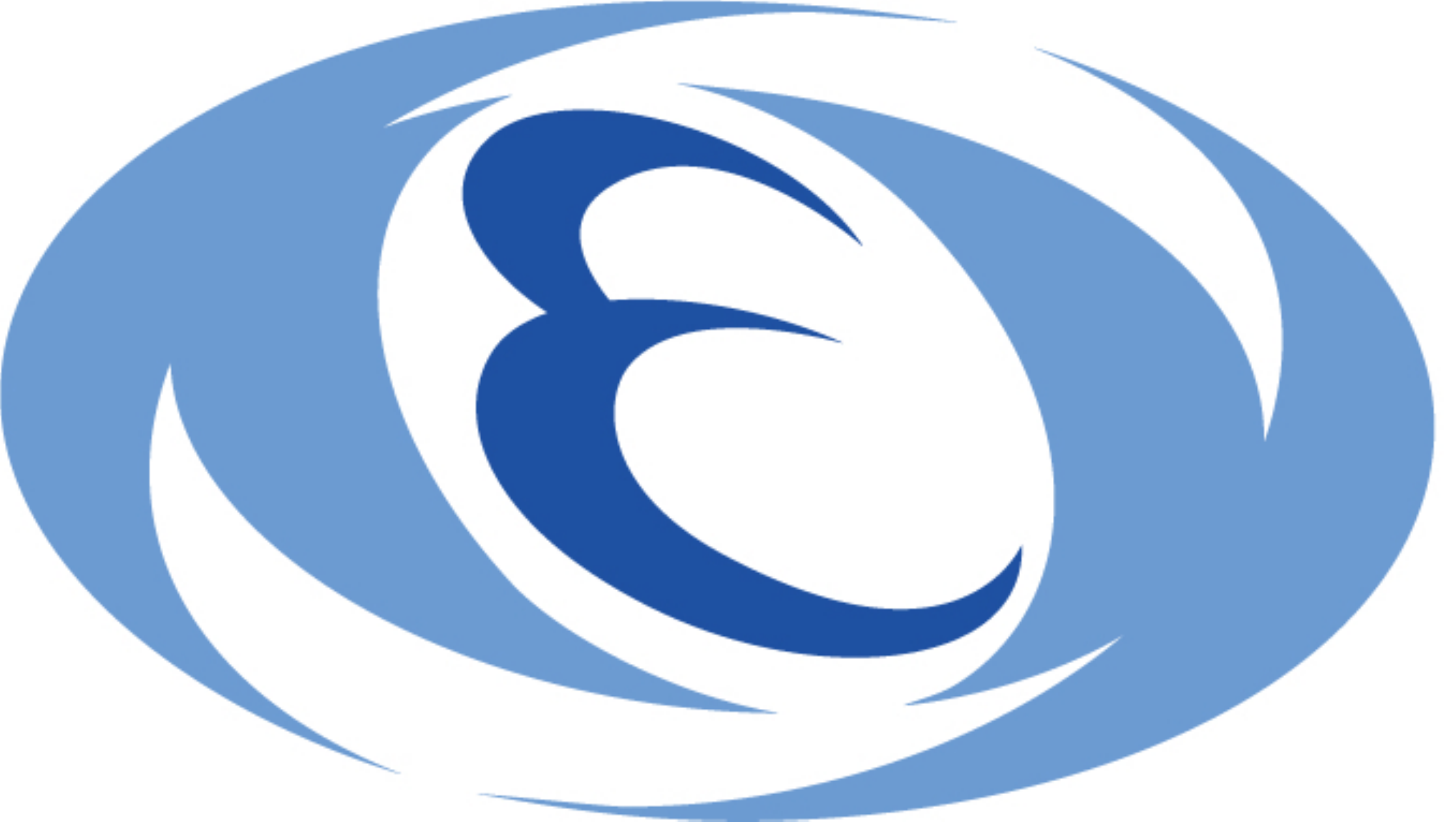}
}}
\newcommand{\kmi}{{
    \includegraphics[height=0.8em]{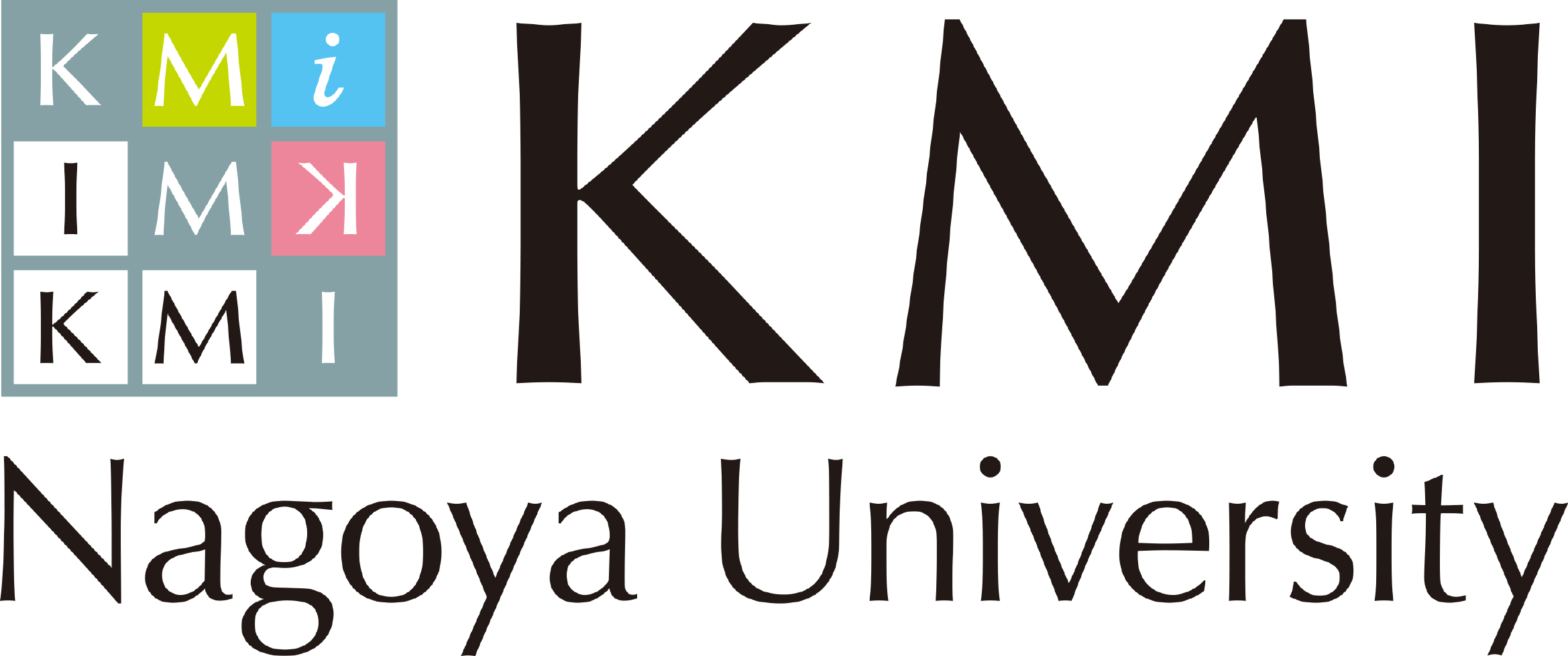}
}}
\newcommand{\nagoya}{{
    \includegraphics[height=0.8em]{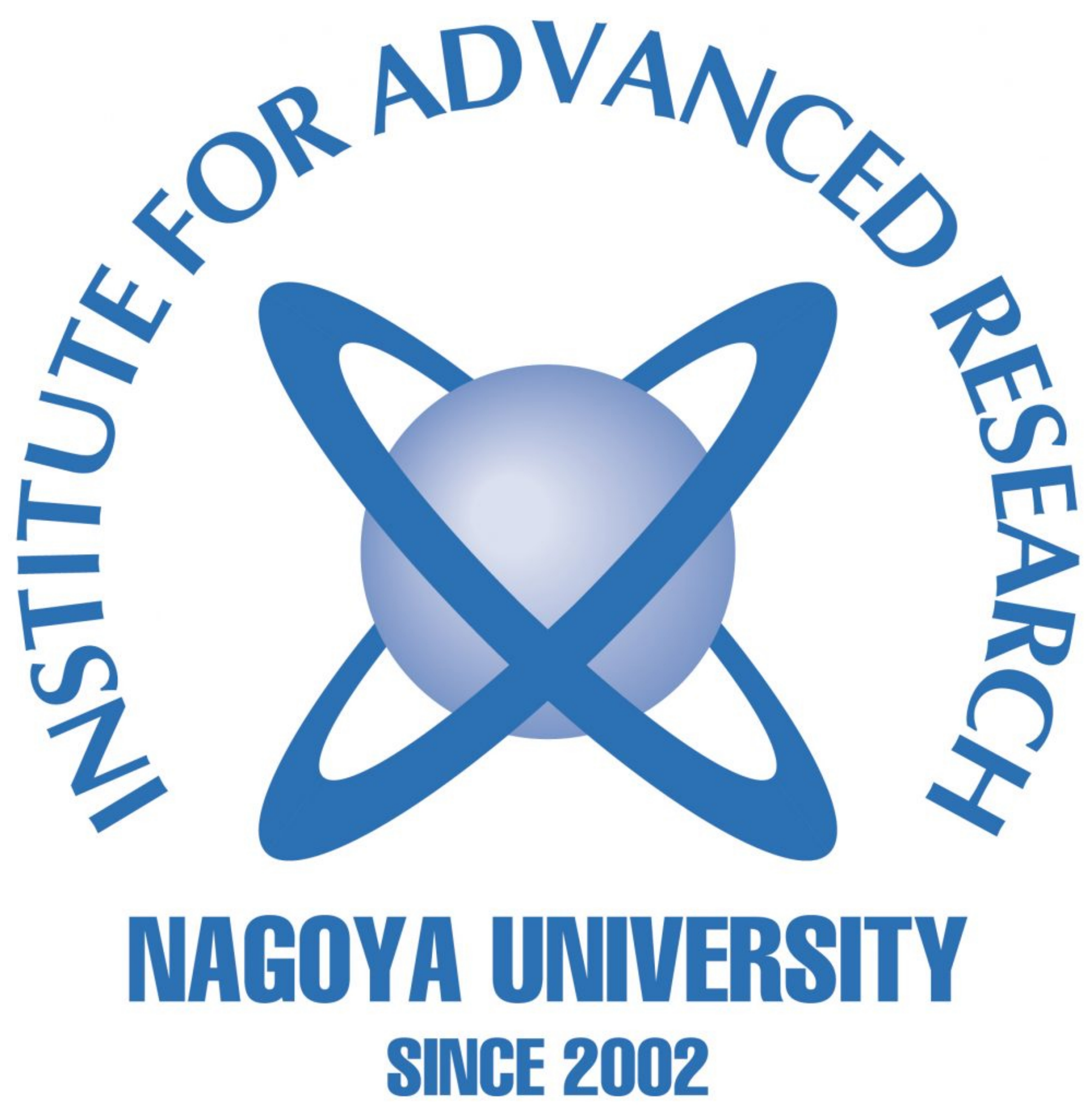}
}}
\newcommand{\sokendai}{{
    \includegraphics[height=0.8em]{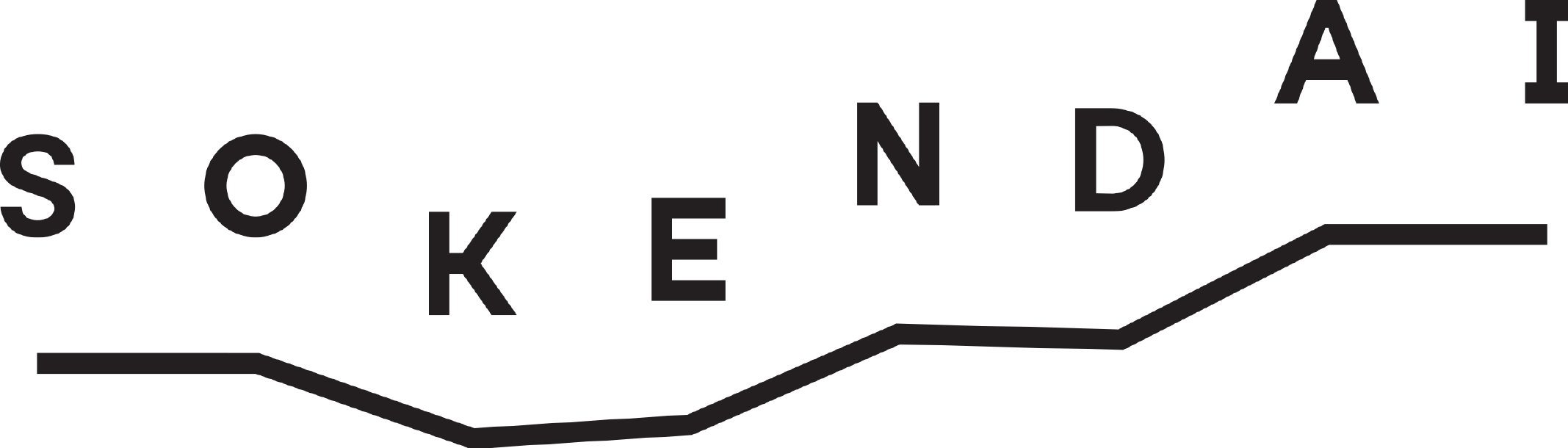}
}}
\newcommand{\tohoku}{{
    \includegraphics[height=0.8em]{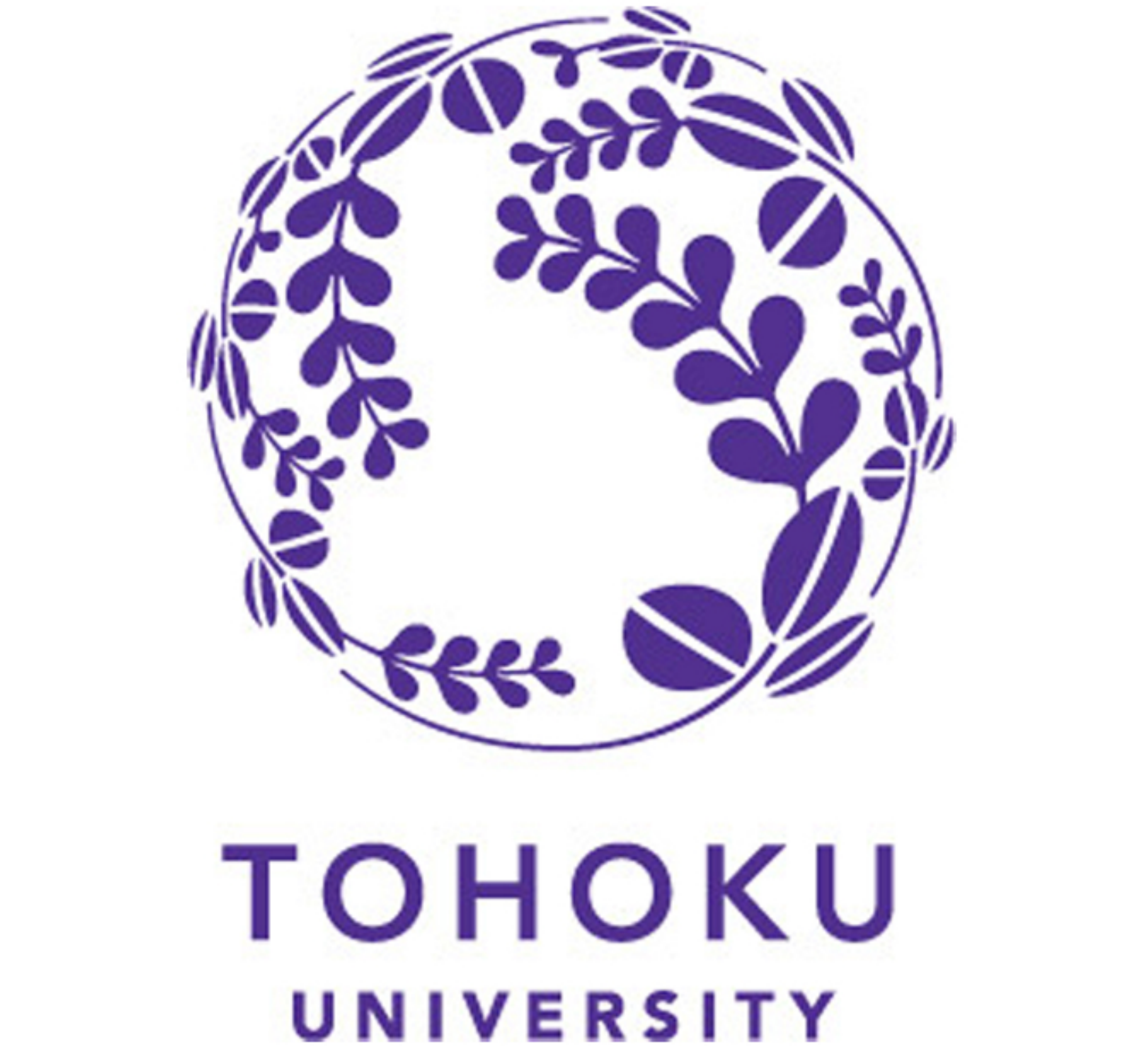}
}}
\newcommand{\fris}{{
    \includegraphics[height=0.8em]{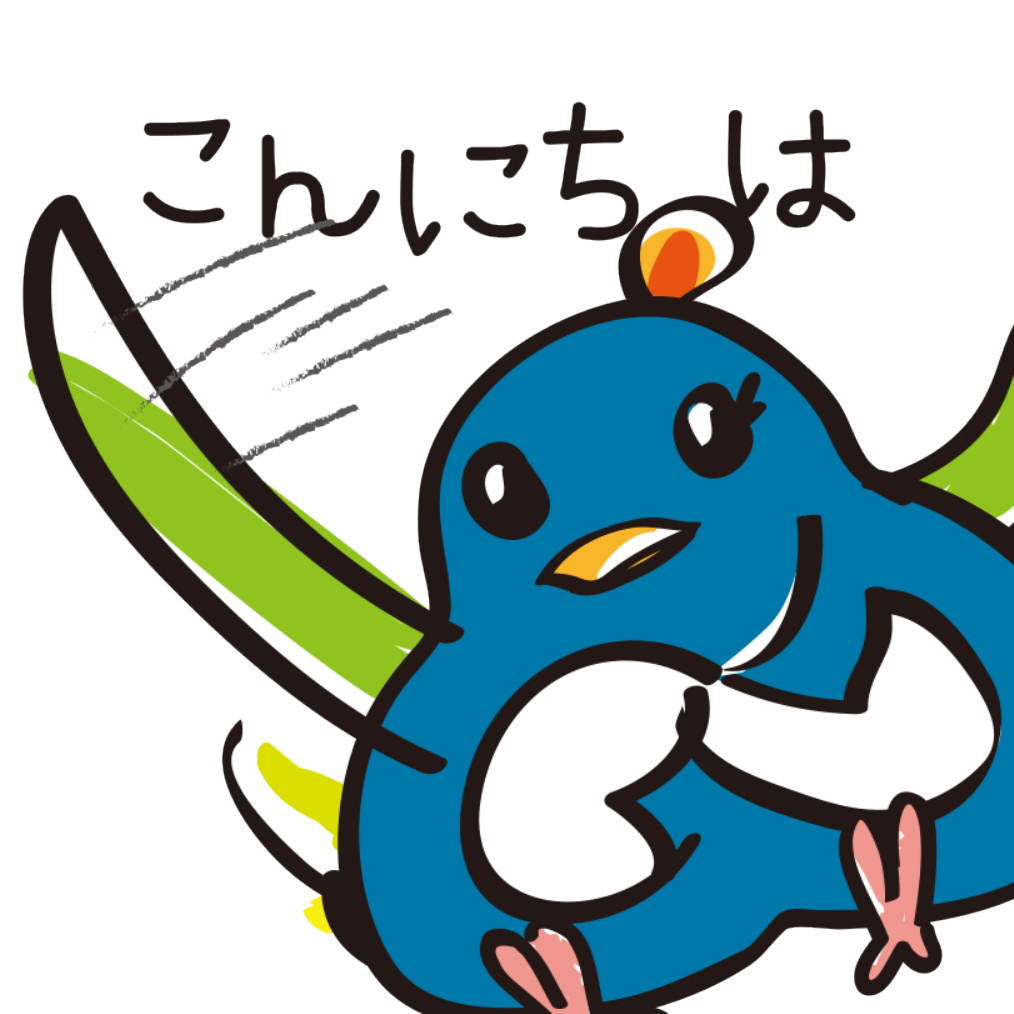}
}}
\newcommand{\harvard}{{
    \includegraphics[height=0.8em]{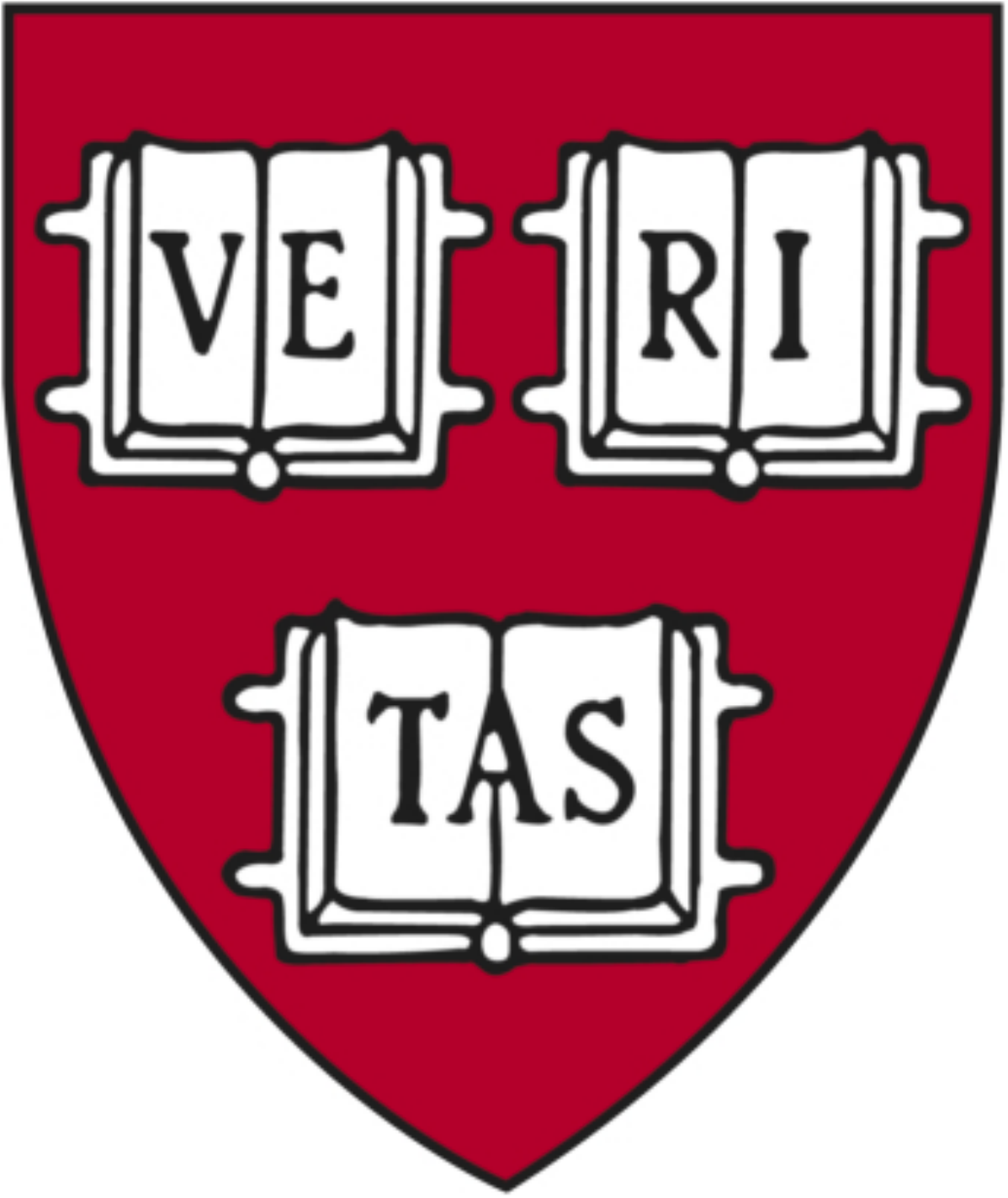}
}}

\title{
Monopole-fermion scattering and varying Fock space
}

\author[\kek,\sokendai,\harvard]{Yuta Hamada,}
\author[\nagoya,\kmi,\kek]{Teppei Kitahara,}
\author[\yitp,\tohoku,\fris]{and Yoshiki Sato}

\affiliation[\kek]{Theory Center, IPNS, High Energy Accelerator Research Organization (KEK), 1-1 Oho, Tsukuba, Ibaraki 305-0801, Japan}
\affiliation[\sokendai]{Graduate University for Advanced Studies (Sokendai), 1-1 Oho, Tsukuba, Ibaraki 305-0801, Japan}
\affiliation[\harvard]{Department of Physics, Harvard University, Cambridge, MA 02138, USA}
\affiliation[\nagoya]{Institute for Advanced Research, Nagoya University, Nagoya, Aichi 464–8601, Japan}
\affiliation[\kmi]{Kobayashi-Maskawa Institute for the Origin of Particles and the Universe, \\
Nagoya University, Nagoya, Aichi  464–8602, Japan}
\affiliation[\yitp]{Center for Gravitational Physics and Quantum Information (CGPQI), \\ Yukawa Institute for Theoretical Physics,
Kyoto University, Sakyo-ku, Kyoto 606-8502, Japan}
\affiliation[\tohoku]{Department of Physics, Tohoku University, Sendai, Miyagi 980-8578, Japan}
\affiliation[\fris]{Frontier Research Institute for Interdisciplinary Sciences,\\ Tohoku University,
Sendai, Miyagi 980-8578, Japan}

\emailAdd{yhamada@post.kek.jp}
\emailAdd{teppeik@kmi.nagoya-u.ac.jp}
\emailAdd{yoshiki.sato@yukawa.kyoto-u.ac.jp}

\abstract{
We propose a four-dimensional interpretation of the outgoing state of the scattering of a massless fermion off a Dirac monopole.
It has been known that such a state has fractional fermion numbers and is necessarily outside the Fock space on top of ordinary perturbative vacuum, 
when more than two flavours of charged Dirac fermions are considered.
In this paper, we point out that the Fock space of the fermions depends on the rotor
degree of freedom of the monopole and changes by a monopole-fermion s-wave scattering.
By uplifting the fermion-rotor system introduced by Polchinski, from two to four dimensions, we argue that the outgoing state can be understood as a state in a different Fock space.
}

\preprint{KEK-TH-2423,~YITP-22-44,~TU-1156}

\begin{document} 
\maketitle
\flushbottom

\section{Introduction}\label{sec:intro}

The magnetic monopole is an important theoretical construct which probes non-perturbative properties or global structures of quantum field theories (QFTs) \cite{Dirac:1931kp,Polchinski:2003bq}, despite the lack of experimental evidence for its existence \cite{MoEDAL:2019ort,MoEDAL:2021vix}. 
Monopoles are also of great importance in understanding our Universe, as any fully-unified theories predict their existence in the form of the 't~Hooft-Polyakov monopole \cite{tHooft:1974kcl,Polyakov:1974ek}.
Absence of magnetic monopoles in our present Universe can be explained by cosmic inflation, due to which the monopole number density could have been diluted \cite{Brout:1977ix,Starobinsky:1980te,Kazanas:1980tx,Sato:1980yn,Guth:1980zm}. 
See the recent review for models and phenomenology of monopoles \cite{Mavromatos:2020gwk}.

Because of their non-perturbative nature, scattering processes involving monopoles can cause peculiar phenomena, such as the Callan-Rubakov effect \cite{Rubakov:1982fp,Callan:1982ah,Callan:1982au,Callan:1982ac}.
It refers to the fact that the monopole in the Grand Unified Theory (GUT) can catalyse proton decay (in the sense that the scattering amplitude is not suppressed by the GUT scale), or more generally fermion-number violating processes, in the limit of vanishing fermion mass.
This is due to the Adler-Bell-Jackiw (ABJ) anomaly of the chiral symmetry (\ie, the baryon number symmetry), whose conservation equation is modified by the instanton number density of the gauge field.
Because of this, the fluctuation of the electric field can conspire with the magnetic field from the monopole to make the instanton number non-zero, causing baryon-number violation.

Baryon-number violations catalysed by monopoles can be seen at the level of the s-wave dynamics.
Let us consider the $U(1)$ gauge theory with $N$ massless Dirac fermions $(\psi_\text{L},\psi_\text{R})$ of charge $+1$ and place a Dirac magnetic monopole (of minimal charge $+1$) at the origin.
As the monopole is a fixed target, solving the Dirac equation in the monopole background amounts to solving the scattering problem \cite{Kazama:1976fm,Rossi:1977im,Callias:1977cc,Goldhaber:1977xw}. 
One can then argue that the only relevant modes are the zero-modes, or the s-wave in our case \cite{Kazama:1976fm,Brennan:2021ucy}.
The zero-modes contain $N$ incoming (outgoing) Dirac fermions, coming from $\psi_\text{L}$ ($\psi_\text{R}$), and their charge assignments under $U(1)_{\text{gauge}}\times SU(N)_{\text{baryon}}$ ($SU(N)_{\text{baryon}}$ is the flavour symmetry) are
\begin{align}
    \begin{tabular}{cccc}
    \bhline{1 pt}
        {4d fermion} & {direction} & $U(1)_{\text{gauge}}$ & $SU(N)_{\text{baryon}}$ \\\hline
        {$\psi_\text{L}$} & incoming & $+1$ & $\square$ \\
        {$\psi_\text{R}$} & outgoing & $+1$ & $\overline{\square}$ \\
    \bhline{1 pt}
\end{tabular}
\end{align}
Obviously, the quantum number under $U(1)_{\text{gauge}}\times  SU(N)_{\text{baryon}}$ needs to be preserved before and after the scattering, if we assume that the monopole is symmetric under the spherical and the $SU(N)_{\text{baryon}}$ symmetry, and that the in- and out-state monopole both have no electric charge.

Having thus prepared, we are ready to see that something peculiar must happen at the monopole core, and that the peculiarity also evolves when one increases $N$.
Let us start from $N=1$ and consider scattering a fermion $\psi_\text{L}$ off a monopole \cite{Kazama:1976fm} --
We immediately see that the symmetry under $U(1)_{\text{gauge}}$ dictates that the out-state can only be $\psi_\text{R}$.
This means that the helicity of the fermion flips before and after the scattering, and that the $U(1)_{\text{baryon}}$ charge is violated by $2$ units.
For $N=2$, if we throw in a fermion $\psi_\text{L}^{a}$, a combination $\epsilon^{ab}\psi_{\text{R},b}$ will come out, where we used the pseudo-reality of the fundamental representation of $SU(2)$. 
This is even more complicated than $N=1$, in that even the flavour changes before and after the scattering in addition to the helicity.

For $N\geq 3$ (in particular $N=4$ which is relevant for the GUT), however, we cannot even find an outgoing state which has the same quantum number as the incoming state so easily.
At least, the outgoing state, hypothetically called the \textit{semiton} \cite{Callan:1982au}, cannot be a combination of one-particle states on top of perturbative vacuum on symmetry grounds.
In order to understand this, it is advantageous to interpret the s-wave part of the monopole-fermion scattering as an effective two-dimensional system of free fermions with a boundary condition, which plays the role of the monopole.
In Refs.~\cite{Callan:1982au,Callan:1983ha,Sen:1983yq,Dawson:1983cm}, it was understood that one should impose the boundary condition on the bosonised fields (or equivalently on the fermion currents), called the dyon boundary condition.
Upon bosonization, the dyon boundary condition simply reflects a soliton into other solitons, and this gives the two-dimensional interpretation of the semiton state \cite{Polchinski:1984uw,Affleck:1993np,Callan:1994ub,Maldacena:1995pq,Smith:2020nuf}.

The remaining question of further interest is the following -- what is this (two-dimensional) solitonic outgoing state in terms of the four-dimensional picture?
It is difficult to find such a candidate in the four-dimensional perspective,
and 
this is sometimes referred to as ``unitarity paradox''.
Many authors have described it in different ways, sometimes in accordance and sometimes in conflict with one another \cite{Polchinski:1984uw,Kitano:2021pwt,Csaki:2021ozp,Brennan:2021ewu}.
We briefly summarise each description in the discussion section.

In this paper, we shed some light on this paradox by pointing out that the bosonisation of the bulk fermion is not necessary in describing the semiton state.
We first give a Lorentzian picture of the boundary scattering of a fermion in the two-dimensional model proposed by Polchinski \cite{Polchinski:1984uw}.
This model couples a theory of bulk free fermions to a rotor degrees of freedom localised at the boundary, and is an s-wave effective theory of the four-dimensional system that we are interested in.
We then do a faithful uplifting of this picture to four dimensions.

This paper is organised as follows.
In Sec.~\ref{sec:paradox}, we consider $U(1)$ gauge theories coupled to massless fermions and introduce Dirac magnetic monopoles.
When we discuss a scattering of a massless fermion off a static monopole, we reduce the 4d model to the 2d model by an s-wave approximation.
After that, we introduce a unitarity paradox in the monopole-fermion scattering.
In Sec.~\ref{sec:Polchinski_model}, we review a two-dimensional phenomenological model introduced by Polchinski for preparation. We point out which boundary conditions we should impose on fermions, a $U(1)$ current, and an $SU(N)$ current.
In Sec.~\ref{sec:proposal}, we propose a solution of the unitarity paradox.
We point out that the Fock space of the fermions depends on the rotor degree of freedom of the monopole and changes by the scattering.
Section~\ref{sec:summary} is devoted to summary and discussion.

\section{Unitarity paradox in monopole-fermion scattering}
\label{sec:paradox}

\subsection{Dirac monopoles in \texorpdfstring{$U(1)$}{U(1)} gauge theories}
\label{sec:def}

Let us consider general $U(1)$ gauge theories in four dimensions and their magnetic line defects, \textit{i.e.}, Dirac monopoles as static probes.
For later reference we couple this to massless fermions, so that we do not have to consider the topological angle.
As is well known, because of the Dirac-Zwanziger quantisation condition \cite{Dirac:1931kp,Zwanziger:1968rs}, the electric and magnetic charges are quantised.
We refer to the minimal electric and magnetic charge as $e$ and $g:= 2\pi/e$, respectively, so that general fields have dyon charge $(\nn e,\,\mm g)$ , where $\nn$ and $\mm$ are integers.
We will hereafter rescale the dyon charges so that they are labelled simply by a set of two integers, $(\nn, \mm)$.

The Dirac-Zwanziger quantisation condition is merely a necessary condition, and it is usually nontrivial if one can consistently add Dirac monopoles (or equivalently magnetic line defects) of any desired charges to the system, even though they are static probes \cite{Smith:2019jnh,Smith:2020rru,Smith:2020nuf,BoyleSmith:2022duw}.
In this work in particular, we are interested in a minimal-charge Dirac monopole, so we choose the matter content so that we surely have such an object.

Based on these, the model we consider in this paper is the $U(1)$ gauge theory coupled to $N$ left-handed Weyl fermions with gauge charge $\nn = \pm 1$ each, denoted as $\psi_{a}^{+}$ and $\psi_{a}^{\prime -}$, with $a=1,\dots, N$, and their complex conjugate as $\bar\psi_{\bar a}^{-}$ and $\bar\psi_{\bar a}^{\prime +}$, respectively (see Table~\ref{tab:charge} for clarity)
\begin{align}
     U(1)+N\times (\text{Weyl with charge $+1$})+N\times (\text{Weyl with charge $-1$})\,.
     \label{eq:setup}
\end{align}
This theory can be UV completed by the following $SU(2)$ gauge theory
\begin{align}
    SU(2) + N\times \text{(fundamental Weyl)} + \text{(adjoint Higgs)}\,.
    \label{eq:su2}
\end{align}
Giving the vacuum expectation value to the adjoint Higgs, we obtain the theory \eqref{eq:setup}, whereas Dirac monopoles are realised from the 't~Hooft-Polyakov monopole inside the theory \eqref{eq:su2}.

As a side comment, we will implicitly assume $N$ to be even from now on unless stated otherwise.
This is because the theory \eqref{eq:su2} is not consistent due to the Witten anomaly unless $N$ is even \cite{Witten:1982fp}.
It has also been argued in Ref.~\cite{McGreevy:2011if,Sato:2022vii} that the existence of a (deconfined) minimal charge monopole for odd $N$ is inconsistent because of the localised Majorana zero mode.
This rules out the possibility of other UV completions which realise \eqref{eq:setup} for odd $N$ (maybe subject to some implicit assumptions made there).

\subsection{Scattering of a fermion off a monopole}\label{sec:scattering}

\subsubsection{Partial-wave decomposition}

Let us consider scattering a charged fermion off a Dirac monopole (with charge $|\nn \mm| \geq 1$) in the model 
\eqref{eq:setup}.
We are interested in the regime where the energy of the scattering is much smaller than the monopole mass so we take its mass to infinity.
This is also equivalent to taking the size of the monopole to be zero for the 't~Hooft-Polyakov monopole of the $SU(2)$ gauge theory \eqref{eq:su2}.

We now take the spherical coordinates
\begin{align}
    \dd s_{\text{4d}}^2  = - \, \dd t^2 + \dd r^2 + r^2 \dd \Omega^2_{\mathbb{S}^2} \,,
\end{align}
with the monopole at the origin of the spatial slice.
The expansion of the fermion fields in terms of spherical harmonics is modified by the background monopole charge, so that the basis becomes the monopole spherical harmonics \cite{Kazama:1976fm} (see, \eg, Ref.~\cite{Shnir:2005vvi} for a pedagogical review).
The total angular momentum $\bm{J}$ is given in Ref.~\cite{Kazama:1976fm} as
\begin{align}
\label{Eq:angular_momentum}
    \bm{J} = \bm{r} \times \left( \bm{p} - \nn e \bm{A}\right) + \frac{1}{2} \bm{\sigma}
    - \frac{\nn \mm }{2} \hat{\bm{r}}\,,
\end{align}
where $\bm{r}$ and $\bm{p} = -\i \partial_{\bm{x}}$ are the position and the three-momentum of the fermion, respectively, while $\bm{A}$ and $\bm{\sigma}$ are the gauge field and the Pauli matrices, respectively.
We have also defined $\hat{\bm{r}} := \bm{r}/r$.
Note that the first term represents the orbital angular momentum, the second the spin of the fermion, and the third the angular momentum of the monopole (coming from the Poynting vector).
Since the total angular momentum \eqref{Eq:angular_momentum}
satisfies the commutation relation,
\begin{align}
    [\bm{J}^2,J_z] = 0\,,
\end{align}
we can diagonalize $\bm{J}^2$ and $J_z$ with eigenvalues $j(j+1)$ and $m$, simultaneously:
the spin spherical harmonics with monopoles satisfies
\begin{align}
    \bm{J}^2 f_{j,m,\nn \mm}^{(\lambda)} (\Omega) &= j(j+1) f_{j,m,\nn \mm}^{(\lambda)} (\Omega) \,, \\ 
    J_z f_{j,m,\nn \mm}^{(\lambda)} (\Omega) &= m f_{j,m,\nn \mm}^{(\lambda)} (\Omega)\,,
\end{align}
with eigenvalues $j$ and $m$ 
\begin{align}
    j &= j_0, \, j_0 +1, \, \dots \,,  \qquad  \text{with} \quad  j_0 := \frac{\left|\nn \mm\right|}{2} - \frac{1}{2}\,, \\
    m &= -j,\, -j+1,\, \dots,\, j-1,\, j\,.
\end{align}
Here, $f_{j,m,\nn \mm}^{(\lambda)} (\Omega)$ is a two-component spinor, which depends on $\nn \mm$ and differs from the usual spin spherical harmonics without a monopole, and the index $\lambda=1,2$ labels the degeneracy of the eigenfunctions if needed.\footnote{To be concrete, $f_{j>j_0,m,\nn \mm}^{(1,2)} (\Omega)$ and $f_{j_0,m,\nn \mm} (\Omega)$ correspond to $\xi_{j>j_0,m}^{(1,2)}$ and $\eta_m$ in Ref.~\cite{Kazama:1976fm}, respectively.}
For $j=j_0$, there is only one solution, and the label $\lambda$ is omitted.
For $j \geq j_0 + 1$, there are two solutions distinguished by the label $\lambda$.

Moreover, the $j=j_0$ mode is an eigenfunction of $\bm{\sigma}\cdot \hat{\bm{r}}$:\footnote{The operator $\bm{\sigma}\cdot \hat{\bm{r}}$ also commutes with $\bm{J}^2$ and $J_z$.}
\begin{align}
    \bm{\sigma}\cdot \hat{\bm{r}} \,f_{j_0,m,\nn \mm} (\Omega) =
    \text{sgn}(\nn \mm )
    f_{j_0,m,\nn \mm} (\Omega) \,.
\end{align}
This condition indicates that 
when $\nn \mm = 1$ $(-1)$, 
the left-handed component must be incoming (outgoing), while the right-handed one is outgoing (incoming), respectively.
In other words, the helicity of the incoming and outgoing waves must be opposite (see Table~\ref{tab:charge}).

For $j\geq j_0+1$, there exists a solution of the massless Dirac equation (with energy $E$) that vanishes at the origin.
The solution of the Dirac equation with energy $E$ is given by \cite{Kazama:1976fm}
\begin{align}
\begin{aligned}
    \psi_{j,E}&=\sum_m\frac{1}{\sqrt{r}}
    \begin{pmatrix}
    a_{j,m,E} \left[\i J_{\mu-\frac{1}{2}}(Er) f_{j,m,\nn \mm}^{(1)} (\Omega) -J_{\mu+\frac{1}{2}}(Er)f_{j,m,\nn \mm}^{(2)} (\Omega) \right] \\[1.em]
    b_{j,m,E} \left[\i J_{\mu-\frac{1}{2}}(Er) f_{j,m,\nn \mm}^{(1)} (\Omega) +J_{\mu+\frac{1}{2}}(Er)f_{j,m,\nn \mm}^{(2)} (\Omega) \right]
    \end{pmatrix}\,,
    \\
    \mu&:=\sqrt{\left(j+\frac{1}{2}\right)^2-\frac{\nn^2\mm^2}{4}}\,,
\label{eq:highj_solution}\end{aligned}
\end{align}
where $J_{\mu \pm \frac{1}{2}} (Er)$ is the Bessel function of order $\mu\pm \frac{1}{2}$, and $a_{j,m,E}, b_{j,m,E}$ are integration constants.\footnote{Here we use the chiral representation while the Dirac representation is adapted in Ref.~\cite{Kazama:1976fm}. These are related by $\gamma^\mu_\text{chiral}=U \gamma^\mu_\text{Dirac} U^\dagger$ where
$U=\frac{1}{\sqrt{2}}\left(\begin{smallmatrix}
    1 & -1\\
    1 & 1
\end{smallmatrix}\right)$.
}
On the other hand, for $j=j_0$ no solution vanishes at $r=0$, the monopole core.
The non-vanishing solution is \cite{Kazama:1976fm} ($\nn \mm >0$ is assumed for simplicity)
\begin{align}
    \psi_{j_0,E}=
    \sum_{m=-j_0}^{j_0}\frac{1}{r}\begin{pmatrix}
    c_{m,E}\, \mathrm{e}^{-\i Er} f_{j_0,m,\nn \mm} (\Omega) \\[0.5em]
    d_{m,E} \,\mathrm{e}^{\mathrm{i}Er} f_{j_0,m,\nn \mm} (\Omega)
    \end{pmatrix}\,,
\label{eq:j0_solution}\end{align}
where $c_{m,E}$ and $d_{m,E}$ are integration constants.
The general solution of 4d Dirac spinor is written as the superposition:
\begin{align}
        \psi(x)=
    \begin{pmatrix}
    \psi^+\\ \bar \psi^{\prime +}
    \end{pmatrix}
    =\int \! \dd E \, \mathrm{e}^{-\i Et} \left(\psi_{j_0,E} + \sum_{j\geq j_0+1}\psi_{j,E}\right) \,. \label{eq:expantion}\end{align}
Note that the left- and right-handed components of the first term in Eq.~\eqref{eq:expantion} have only incoming and outgoing modes, respectively:
\begin{align}
    \int \! \dd E \, \mathrm{e}^{-\i Et} \psi_{j_0,E} =: \sum_m
    \frac{1}{r}\begin{pmatrix}
    \chi^+_{\text{in},m}(t+r) f_{j_0,m,\nn \mm} (\Omega) \\
    \chi^+_{\text{out},m}(t-r) f_{j_0,m,\nn \mm} (\Omega)
    \end{pmatrix}  \,.
\end{align}
This is in contrast to the $j\geq j_0+1$ modes, where left/right-handed component has both incoming and outgoing modes (since the Bessel function behaves as $\sin(Er)/\sqrt{E r}$ or $\cos(Er)/\sqrt{E r}$ for large $E r$).
This implies that, in the context of the unitarity paradox of the monopole-fermion scattering, only the first term in Eq.~\eqref{eq:expantion} needs to be kept (see Sec.~\ref{sec:2d}).
In the following, we call this truncation as s-wave approximation.
The generalisation to the multiple flavour case is straightforward:
\begin{align}
    \begin{pmatrix}
    \psi_a^+\\ \bar \psi^{\prime +}_{\bar{a}}
    \end{pmatrix}
    =\sum_m\frac{1}{r}\begin{pmatrix}
    \chi^+_{\text{in},a,m}(t+r) f_{j_0,m,\nn \mm} (\Omega) \\
    \chi^+_{\text{out},\bar{a},m}(t-r) f_{j_0,m,\nn \mm} (\Omega)
    \end{pmatrix}
    + (j\geq j_0+1 \,\text{modes}) \,.
\end{align}

Near the core of the monopole $r\sim0$, the solution behaves as 
\begin{align}
    \psi_{j,E}\propto r^{\mu-1}\,,
\end{align}
from Eqs.~\eqref{eq:highj_solution} and \eqref{eq:j0_solution}.
Thus, if we interpret $\psi_{j,E}$ as a wave function of the fermion with energy $E$ and angular momentum $j$, the probability that the fermion exists near $r=0$ is\footnote{Note that the solution we discarded in Eq.~\eqref{eq:highj_solution} is $\psi_{j,E}\sim r^{-\mu-1}$ for $r\sim0$. This leads to the divergent probability $r^2 |\psi_{j,E}|^2\sim r^{-2\mu}$, which is presumably physically unacceptable.} 
\begin{align}
    r^2 \left| \psi_{j,E} \right|^2 \propto r^{2\mu} \,, 
\end{align}
where $r^2$ comes from a volume factor.
This means that only the $j=j_0$ mode, which leads to $\mu=0$, reaches the monopole core. Since the $U(1)$ monopole solution is singular at $r=0$, we have to impose the boundary condition in order to define the theory.
Next, we will discuss the issues of the boundary condition.

\subsubsection{Effective two-dimensional description}\label{sec:2d}

\begin{table}[t]
\centering
\renewcommand{\arraystretch}{1.5}
\rowcolors{2}{gray!15}{white}
\addtolength{\tabcolsep}{3pt} 
\begin{tabular}{ccccc}
   \bhline{1 pt}
          2d Weyl&  4d Weyl & $\nn  $ & $\bm{\sigma} \cdot \hat{\bm{r}}$ (4d)         & helicity (4d)    \\
          \hline
$\chi_{\text{in},a}^+$  & $\psi^+_a$ &  $1$ & $1$          &  $-1/2$  (left-handed)   \\
$\chi_{\text{in},\bar{a}}^-$  & $\bar \psi^-_{\bar{a}}$ & $-1$ & $-1$          &    $1/2$ (right-handed) \\
$\chi_{\text{out},a}^-$ & $\psi^{\prime -}_a$ &  $-1$ & $-1$         &  $-1/2$  (left-handed) \\
$\chi_{\text{out},\bar{a}}^+$  &  $\bar \psi^{\prime +}_{\bar{a}}$ & $1$ & $1$          &   $1/2$  (right-handed) \\
\bhline{1 pt}
\end{tabular}
\addtolength{\tabcolsep}{-5pt} 
\caption{
Correspondence between 4d and 2d fermions as a result of the partial wave decomposition, in the presence of the minimal-charge Dirac monopole ($\mm=1$) and in the lowest partial wave sector.
4d and 2d fields are respectively denoted using $\psi$ and $\chi$.
For both $\psi$ and $\chi$, $\pm$ denotes the $U(1)$ charge of the fermions.
In addition, the 4d fields with or without bar means that the chirality of the field is right- or left-handed, while for the 2d fields in/out means that the fields are incoming/outgoing with respect to the boundary at $r=0$.
Flavour indices of the fermions are indicated by $a {, \bar{a}}=1,\dots, N$.
We also take the particles in the white entries as a particle, and grey as their antiparticles.
Note that the gauge group $U(1)$ turns in to the $U(1)_A$ symmetry in the two-dimensional description because of this.
}
\label{tab:charge}
\end{table}

Each angular momentum mode of the four-dimensional fermions can be thought of as a two-dimensional fermion \cite{Rubakov:1982fp,Brennan:2021ewu}. 
This effectively reduces the original system to a $1+1$ dimensional one, with spacetime coordinates being $(t,r>0)$,
where the boundary condition at $r=0$ plays the role of the monopole.
We assume that this boundary condition does not mix different angular momentum modes, which is tantamount to assuming that the monopole is spherically symmetric.
We also assume that this effective 2d system as well as its boundary condition is conformal in the IR limit.

As we can see from the discussion above, the $j \geq j_0 +1 $ partial waves of one Weyl fermion come in a pair of incoming and outgoing modes, 
while the $j=j_0$ wave of one Weyl fermion is chiral and only contains incoming or outgoing mode by virtue of the Atiyah-Singer index theorem \cite{BoyleSmith:2022duw}.
For the $j\geq j_0 +1$ partial waves, the incoming and the outgoing modes are coupled \textit{via} the position-dependent mass term.
As the profile of the mass becomes a delta function in the limit of infinite monopole mass \cite{Brennan:2021ewu}, we can simply impose the Dirichlet boundary condition for those modes.\footnote{This position-dependent mass was computed from studying the 't~Hooft-Polyakov monopole and taking the infinite mass limit. Even though this argument is not purely in terms of the original $U(1)$ gauge theory, we will adopt this result for our case as well.}
We will not discuss these higher angular momentum modes anymore.

On the other hand, for the $j=j_0$ partial wave, the effective two-dimensional modes are chiral as discussed. 
In other words, $\psi^{+}$ and $\bar{\psi}^{-}$ create incoming modes only in the two-dimensional effective description, denoted as $\chi_{\mathrm{in}}^{+}$ and $\chi_{ \mathrm{in}}^{-}$, respectively.
We summarised the correspondence between 4d and 2d modes in the lowest spin partial wave sector, in Table~\ref{tab:charge}.
Since we do not have a natural boundary condition for those chiral modes, understanding their boundary condition amounts to understanding the monopole-fermion scattering in question.

Such boundary conditions can be characterised using the symmetries that they preserve.
It is usually believed that when a certain symmetry is non-anomalous, a boundary condition exists which is invariant under it.\footnote{This statement of existence is only a folklore, and only the converse is proven in Ref.~\cite{Thorngren:2020yht}. See Ref.~\cite{Hellerman:2021fla} for a list of statements related to anomaly and boundary. See also Refs.~\cite{Smith:2019jnh,Smith:2020nuf,Smith:2020rru} for concrete examples where this is true. Note also that the relation between anomaly and the existence of general codimension-$p$ defects is not known yet.}
From now on, we choose such an anomaly-free symmetry to be $U(1)_A\times SU(N)_V$, in the 2d language, which is $U(1)_{\text{gauge}}\times SU(N)_{\text{baryon}}$ symmetry in the 4d language;
\begin{align}
    \begin{tabular}{cccc}
    \bhline{1 pt}
        {2d Weyl} & {4d Weyl} & $U(1)_A$ & $SU(N)_V$ \\\hline
        {$\chi_{\text{in},a}^+$} & {$\psi^+_a$} & $+1$ & $\square$ \\
        {$\chi_{\text{out},a}^-$} & {$\psi^{\prime -}_a$} & $-1$ & $\square$ \\
    \bhline{1 pt}
\end{tabular}
\end{align}
Note that we could have taken $SU(N)_A$ instead of $SU(N)_V$, but we do not study this possibility since this is not of interest in view of the  't~Hooft-Polyakov monopoles of $SU(2)$ gauge theory -- $SU(N)_A$ symmetry is only emergent after Higgsing of the non-diagonal gauge fields.

\subsection{Boundary condition of the effective two-dimensional theory}
\label{sec:boundary_conditions}

\subsubsection{Free boundary condition}

Free boundary condition is one of the simplest boundary conditions of the two-dimensional effective theory.
This boundary condition is usually placed at $r=R$, far away from the monopole, and is written as
\begin{align}
\mathtt{A}[\theta]: \qquad 
    \left. \chi_{\mathrm{in},a}^{+} \right|_{r=R} = \left. \mathrm{e}^{\i \theta} \delta_{a,\bar{a}} \chi_{\mathrm{out},\bar{a}}^{+} \right|_{r=R} \,,
    \label{eq:free}
\end{align}
and its complex conjugate, where $\theta$ is some arbitrary parameter.
Different $\theta$ are connected \textit{via} the $U(1)_V$ rotations.
This boundary condition preserves the $U(1)_A$ symmetry but not the $SU(N)_V$ flavour symmetry.
Note that $N$ denotes the number of $SU(2)$ Weyl fermion doublet and is also the number of $U(1)$ Dirac fermion.

In the four-dimensional language, this boundary condition can be obtained from a position-dependent mass term for the Weyl fermion,
\begin{align}
    M(r)\psi^+_{a}{\psi}_{{a}}^{\prime -} \,,
\end{align}
where $M(r)$ is zero when $r<R$, while taken to be large when $r>R$.
{The massless fermion can not penetrate the region of $r > R$, which is consistent with Eq.~\eqref{eq:free}.}
The arbitrary parameter $\theta$ in Eq.~\eqref{eq:free} is encoded in the phase of $M(r)$.
Hereafter, we denote this boundary condition as $\mathtt{A}[\theta]$-boundary.

\subsubsection{Dyon boundary conditions}

Dyon boundary conditions are the boundary conditions that are invariant under $U(1)_A \times SU(N)_V$.
As discussed, this class of boundary conditions corresponds to the magnetic line defect in the four-dimensional gauge theory of interest.
It is known that such boundary conditions satisfy the following relation \cite{Polchinski:1984uw,Maldacena:1995pq,Affleck:1993np},
\begin{align}
\mathtt{D}[\theta]: \qquad 
    \left. R_{ab}J_{\mathrm{in},b}\right|_{r=0} = \left. J_{\mathrm{out},a} \right|_{r=0}\,, \qquad  R_{ab} :=  \delta_{ab}-\frac{2}{N} \,, 
    \label{dyon_bc}
\end{align}
where
\begin{align}
    J_{\mathrm{in},a} := \sum_{\bar{a}=1}^N \chi^{+}_{\mathrm{in},a}\chi^{-}_{\mathrm{in},\bar{a}} \delta_{a,\bar{a}} \quad \text{and} \quad  J_{\mathrm{out},a} := \sum_{\bar{a}=1}^N \chi^{-}_{\mathrm{out},a}\chi^{+}_{\mathrm{out},\bar{a}} \delta_{a,\bar{a}} \,.
\end{align}
The reason for the free parameter $\theta$ will be explained in the next paragraph.
Note that the expression above only makes manifest the $U(1)^N\subset U(1)_A \times SU(N)_V$ symmetry at the boundary.

As indicated in the notation, dyon boundary conditions are secretly parameterised by a free parameter $\theta$.
This is analogous to $\mathtt{A}[\theta]$, as different $\theta$ are connected \textit{via} $U(1)_V$ rotations.
We usually identify $\theta$ as the topological angle of the $U(1)$ gauge theory, as discussed by Refs.~\cite{Polchinski:1984uw,Affleck:1993np}.
The boundary condition corresponding to the Dirac monopole is known to be $\mathtt{D}[\theta]$.

For $N=1,2$, we can impose linear dyon boundary conditions not on currents but fermions as we will see later.
As proved in Refs.~\cite{Polchinski:1984uw,Callan:1983ed}, these linear boundary conditions are the same as $\mathtt{D}[\theta]$ \eqref{dyon_bc}.

For the theory of our interest, the boundary conditions that we need are already constructed in the literature \cite{Callan:1983ed,Callan:1982au,Callan:1983ha,Sen:1983yq,Dawson:1983cm}. 
We review them for $N=2$ and for $N\geq 4$, separately.
We will also review the case with $N=1$ for pedagogical reasons, even though we have promised, in Sec.~\ref{sec:def}, to look only at the case where $N$ is even, because of the Witten anomaly.

\subsubsection{Unitarity paradox}

From now on, we consider a $\nn \mm =1$ case for simplicity.
In this case, $\chi_{\text{in}}^+$ and $\chi_{\text{out}}^+$ have no angular momentum index because of $j_0 =0$.

For $N=1,2$, we will see that a final state exists.
In contrast to the $N=1,2$ cases, a final state is missing for the $N \geq 4$ cases.

\paragraph{$N=1$ case}

As a warm-up, we first consider the $N=1$ case.
Even though the $N=1$ case suffers from the Witten anomaly as noted in the previous section, this setup reveals the strange feature of the scattering of the fermion off the monopole.

In the s-wave approximation, 
when an incoming fermion is $\chi_\text{in}^+$, 
the conservation of the angular momentum implies that a candidate of an outgoing fermion is $\chi_\text{out}^\pm$, and 
the conservation of the $U(1)$ charge gives a further restriction.
Then, the scattering process is denoted in the two-dimensional perspective as 
\begin{align}
    M + \chi_\text{in}^+ \to M + \chi_\text{out}^+\,,
\label{eq:N=1}\end{align}
where $M$ is the Dirac monopole.
For the $N=1$ case, the dyon boundary condition can be written as
\begin{align}
        \left. \chi_{\text{in}}^+ \right|_{r=0} = \left. \mathrm{e}^{\i \varphi}  \chi_{\text{out}}^+ \right|_{r=0} \,.
        \label{eq:boundaryN1}
    \end{align}
The phase factor $\mathrm{e}^{\i \varphi}$ is not invariant under the chiral rotation of the fermion, while
the combination of invariant angles is
\begin{align}
  \overline{\vartheta} =  \varphi-\vartheta\,,
\label{eq:invariant_angle}\end{align}
where $\vartheta$ is the coefficient of $U(1)$ $F_{\mu\nu}\tilde{F}^{\mu\nu}$ term in the parity-violating  $U(1)$ gauge theory.\footnote{See Ref.~\cite{Hamada:2022bzn} for the recent analysis of the $N=1$ case with the time-dependent topological angle.}

In the four-dimensional perspective,
the scattering process is denoted as 
\begin{align}
    M + \psi^+  \to M + \bar\psi^{\prime +} \,,
\end{align}
and it implies that the fermion helicity flips \cite{Kazama:1976fm}.
Therefore, the angle $\varphi$ in Eq.~\eqref{eq:boundaryN1} is a  parity and $CP$ violating parameter.

\paragraph{$N=2$ case}

Next, let us consider the $N=2$ case and the situation where an incoming fermion is $\chi_{\text{in},1}^+$.
From the conservation of the angular momentum,
the fermion helicity flips.
In addition to this flip, the flavour index of the fermion must change since the dyon boundary condition at the core can be written as \cite{Callan:1983ed,Polchinski:1984uw} 
\begin{align}
    \left. \chi_{\text{in},a}^+ \right|_{r=0} = \left. \i\, \mathrm{e}^{\i \varphi} \epsilon_{a\bar{b}} \,\chi_{\text{out},\bar{b}}^+ \right|_{r=0} \,.
\label{eq:boundaryN2}\end{align}
Again, the angle $\varphi$ is a parity violating, and the invariant angle is given by \eqref{eq:invariant_angle}. 
This observation implies that the scattering of $\chi_{\text{in},1}^+$ off the monopole $M$ is 
\begin{align}
    M + \chi_{\text{in},1}^+ \to M + \chi_{\text{out},\bar{2}}^+\,,
\label{eq:N=2}\end{align}
where we add a bar for particles with index $\bar{a}$.
This process is translated to
\begin{align}
    M + \psi_{1}^+  \to M + \bar \psi_{2}^{\prime +} \,,
\end{align}
in the four-dimensional perspective.

\paragraph{$N \geq 4$ cases}

Finally, let us consider a scattering of the $N \geq 4$ cases,
\begin{align}
M + \chi_{\text{in},a}^+ \to M + X \,,
\label{???}
\end{align}
where $X$ denotes the unknown outgoing fermion.
When the incoming fermion is $\chi_{\text{in},1}^+$ for example, the candidate of the final state $X$ should be $\chi_{\text{out},\bar{a}}^+$ (or their superposition) due to the conservation of the angular momentum and the $U(1)$ charge.
However, we cannot maintain the flavour symmetry at the core for the $N \geq 4$ case by imposing a linear dyon boundary condition in contrast to the $N=1,2$ cases \cite{Callan:1983ed,Dawson:1983cm}.
This suggests that the final state does not exist.
There are two problems:
 \begin{itemize}
    \item[(i)] a missing of the final state in the two-dimensional perspective, and
    \item[(ii)] if the problem (i) is solved, the interpretation of the final state in the four-dimensional perspective.
\end{itemize}
The problem (i) was solved by bosonisation and imposing $\mathtt{D}[\theta]$-boundary \eqref{dyon_bc} for the fermions in Refs.~\cite{Callan:1983ha,Sen:1983yq,Polchinski:1984uw,Maldacena:1995pq}. It turned out that the final state has the fractional fermion number, which is called the \textit{semiton} state. 
On the other hand, the problem (ii) is in ongoing debate. See Refs.~\cite{Kitano:2021pwt,Csaki:2021ozp,Brennan:2021ewu} for recent proposals.
In the following sections, we review the solution of problem (i) following Ref.~\cite{Polchinski:1984uw}, 
and then we provide a possible solution of the problem (ii). The relation with other proposals is discussed in Sec.~\ref{sec:comments}.
The problem (ii) is known as the unitarity paradox of the monopole-fermion scattering.

\section{Two-dimensional phenomenological model}
\label{sec:Polchinski_model}

In this section, we review a two-dimensional phenomenological model introduced by Polchinski \cite{Polchinski:1984uw} (see also Ref.~\cite{Affleck:1993np}) and derive boundary conditions for fermions and currents,
as preliminaries to the next section.
To compare with the 2d model in Sec.~\ref{sec:paradox}, a rotor degree of freedom, which serves as that of monopole, is introduced  near the origin.
This model can be derived from the Georgi-Glashow model, which includes an 't~Hooft-Polyakov monopole, and reduces to the 2d model in Sec.~\ref{sec:paradox} in a low energy limit.
Therefore, this model is suitable for our purpose.

\subsection{Model}

In a scattering process, we have left-moving fermions $\chi_\text{in}^\pm$ and right-moving fermions $\chi_\text{out}^\pm$.
Since $\chi_\text{in}^\pm$ and $\chi_\text{out}^\pm$ do not interact with each other, it is convenient to extend the spacetime and introduce right-moving fermions,\footnote{This trick, called the unfolding, is not always valid even though the opposite procedure, folding, is always possible. 
}
\begin{align}
\label{eq:unfolding}
    \chi_{\bar{a}} (t,x) = 
    \begin{dcases}
    \chi_{\text{out},\bar{a}}^+ (t-x)\,, & (x >0)\,, \\
    \chi_{\text{in},\bar{a}}^- (t-x)\,, & (x <0)\,,
    \end{dcases} \qquad 
    \chi_a^\dagger (t,x) = 
    \begin{dcases}
    \chi_{\text{out},a}^- (t-x)\,,  & (x >0)\,, \\
    \chi_{\text{in},a}^+ (t-x)\,,  & (x <0) \,.
    \end{dcases}
\end{align}
Here, the awkward suffix $\bar{a}$ appears in $\chi_{\bar{a}}$ because we use right-moving fermions.
In this construction, the electric charge of $\chi_{\bar{a}}$  changes at $x=0$.
We study the Hamiltonian which describes $N$ flavour of a right-moving fermion coupled to a time-dependent rotor degree of freedom $\alpha(t)$ 
\begin{align}
    H = \sum_{a,\bar{a}=1}^{N} \int_{-\infty}^\infty \dd x \, \chi_a^\dagger (x) \left( - \i  \frac{\partial}{\partial x} -  \alpha f(x) \right) \chi_{\bar{a}} (x) \delta_{a,\bar{a}} + \frac{\Pi^2}{2I} \,.
\end{align}
The first term describes fermions in the presence of the vector potential $eA_1 = \alpha (t) f(x)$.\footnote{This relation can be derived from a low energy description of the Georgi-Glashow model.}
The second term is a kinetic term of the rotor; 
$\Pi (t)$ is a conjugate momentum of $\alpha (t)$, which satisfies the commutation relation
\begin{align}
    [\alpha, \Pi] = \i \,, 
\end{align}
and $I$ is the moment of inertia of the rotor.
The monopole-fermion interaction is localised by the function $f(x)$, which is an even function and vanishes outside of the core of dyon,
\begin{align}
    f(x) = 0\,, \quad \text{for} \quad |x| > r_0 \,,
\end{align}
where $r_0$ is the dyon core size, and we normalise  
\begin{align}
    \int_{-\infty}^\infty \! \dd x\, f(x)  = 1 \,.
\end{align}

The conserved electric charge is\footnote{We follow the notation in Ref.~\cite{Affleck:1993np} which is twice of the notation in Ref.~\cite{Polchinski:1984uw}.}
\begin{align}
\label{eq:charge_op}
    Q = 2 \left( \Pi + \int_{-\infty}^\infty \dd x \, q(x) J (x) \right) \,,
\end{align}
where the current is introduced as 
\begin{align}
\label{eq:U(1)current}
    J(x) := \sum_{a,\bar{a}=1}^N J_{a\bar{a}} (x) \delta_{a,\bar{a}} \,, \qquad  J_{a\bar{a}}  := \chi_a^{\dagger} \chi_{\bar{a}} \,,
\end{align}
and the function $q(x)$ is introduced as follows 
\begin{align}
    \frac{\dd q}{\dd x} = -f \,, \quad \text{with} \quad q(x) = \begin{dcases}
    -\frac{1}{2}\,, & (x>r_0)\,, \\
    \phantom{-} \frac{1}{2}\,, & (x<-r_0) \,.
    \end{dcases}
\end{align}
The charge operator \eqref{eq:charge_op} does not commute with the Hamiltonian $H$ due to the anomalous Schwinger term in the current commutator
\begin{align}
    [J(x), J(y)] = - \frac{\i N}{2\pi} \partial_x \delta (x-y) \,.
\end{align}
In other words, the charge $Q$ is conserved classically, but is not conserved quantum mechanically.
This problem can be solved by adding an additional term to the Hamiltonian
\begin{align}
    H_\text{new} = H +\frac{1}{2}C\alpha^2 \,, \qquad C := \frac{N}{2\pi} \int_{-\infty}^\infty \dd x \, f^2 (x) \,.
\label{eq:Polchinski}
\end{align}
The charge $Q$ commutes with the new Hamiltonian,
\begin{align}
    [Q,H_\text{new}] = 0 \,, 
\end{align}
and $Q$ is conserved.
In the following, we study the new Hamiltonian $H_\text{new}$ instead of $H$.

\subsection{Boundary condition}

In this section, we derive boundary conditions for $\chi_{\bar{a}}$, a $U(1)$ current,
and an $SU(N)$ current 
by assuming that incoming particles have low energy $E \ll 1/I$. 
After that, we obtain a change of fermion numbers in a scattering process.

Let us first derive a boundary condition for $\chi_{\bar{a}}$ imposed at the core.
The equation of motion for $\chi_{\bar{a}}$ is given by
\begin{align}
\label{eq:EOM_for_chi}
    \frac{\partial}{\partial t} \chi_{\bar{a}} (t,x) = - \frac{\partial}{\partial x} \chi_{\bar{a}} (t,x) + \i \alpha (t) f(x) \chi_{\bar{a}} (t,x) \,.
 \end{align}
From the low energy observer, the detail description of the monopole core is not needed.
One only needs the boundary condition of the fermions at $x= \pm r_0$. 
With the rotor degree of freedom $\alpha$, the boundary condition preserving $U(1)$ and flavour symmetries is written as\footnote{Since $[Q, \mathrm{e}^{\i \alpha}] = 2 \mathrm{e}^{\i \alpha}$, the boundary condition \eqref{eq:bc_at_origin} preserves the $U(1)$ charge \eqref{eq:charge_op}.}
\begin{align}
    \left. \chi_{\bar{a}} \right|_{x=r_0} = \left. \mathrm{e}^{\i \alpha(t)}\chi_{\bar{a}} \right|_{x=-r_0} \,, 
\label{eq:bc_at_origin}
\end{align}
where a small core approximation ($r_0 E\ll1$ with $E$ being the energy of $\chi$) is used.
The fermion-rotor interaction term in the Hamiltonian \eqref{eq:Polchinski} is chosen so that this boundary condition is realised.\footnote{\label{footnote12}We comment on the $2\pi$ periodicity of $\alpha$ \cite{Polchinski:1984uw}.
This is not symmetry of the Hamiltonian \eqref{eq:Polchinski}.
However, with $F$ being the fermion number and $Q$ being Eq.~\eqref{eq:charge_op}, an operator
\begin{align}
     \Omega=(-1)^F \mathrm{e}^{2\pi \i Q} 
\end{align}
commutes with the Hamiltonian \eqref{eq:Polchinski}, and increases $\alpha$ by $2\pi$.
Therefore, it is possible to diagonalize both \eqref{eq:Polchinski} and $Q$, where the $2\pi$ periodicity is manifest.
In this sense, the variable $\alpha$ looks similar to the topological angle, although the precise relation is not clear.
}

Next, we drive boundary conditions for the $U(1)$ current \eqref{eq:U(1)current}.
In standard electroweak theory and GUT theory, the baryon number current is not a gauge-invariant conserved current.
We can define a current $J$ which is conserved but not gauge-invariant, or a current $\tilde{J}$ which is gauge-invariant but not conserved.

In the Polchinski's model, $J$ is a conserved current,
\begin{align}
    \frac{\partial J_{a\bar{a}}}{\partial t} = - \frac{\partial J_{a\bar{a}}}{\partial x} +  \frac{\alpha}{2\pi} \frac{\dd f}{\dd x} \,,
\end{align}
up to the anomaly.
However, $J$ is not gauge-invariant as it does not commute with $Q$
\begin{align}
    [J_{a\bar{a}},Q] = \frac{\i}{\pi} f \delta_{a,\bar{a}} \,.
\end{align}
We can construct an improved current 
\begin{align}
    \tilde{J}  = \sum_{a,{\bar{a}}=1}^N \tilde{J}_{a{\bar{a}}} \delta_{a,{\bar{a}}} \,, \qquad 
    \tilde{J}_{a{\bar{a}}} (t,x) := J_{a{\bar{a}}}(t,x) - \frac{1}{2\pi} \alpha (t) f(x) \delta_{a,{\bar{a}}} \,,
\end{align}
such that the improved current is gauge-invariant, \ie, it commutes with $Q$, $[\tilde{J},Q] = 0$.
However, $\tilde{J}$ does not commute with $H_{\text{new}}$, $[\tilde{J},H_{\text{new}}] \neq 0$. 
The conservation law for $\tilde{J}_{a\bar{a}}$ becomes
\begin{align}
\label{eq:conservation_law_tildeJ}
    \frac{\partial \tilde{J}_{a\bar{a}}}{\partial t} = - \frac{\partial \tilde{J}_{a\bar{a}}}{\partial x} - \frac{f}{2\pi I} \Pi \delta_{a,\bar{a}} \,.
\end{align}
The equation of motion for $\Pi$, which is obtained from $H_\text{new}$, can be written using $\tilde{J}$,
\begin{align}
 \label{eq:eom_for_Pi}
    \frac{\dd \Pi}{\dd t} = \int_{-\infty}^\infty \dd x \, f(x) \tilde{J} (x) \,.
\end{align}
From Eq.~\eqref{eq:conservation_law_tildeJ}, we obtain 
\begin{align}
\begin{aligned}
    \tilde{J}_{a\bar{a}} (t,x) & = \tilde{J}_{a\bar{a}} (t-x-r_0,-r_0) - \frac{\delta_{a,\bar{a}}}{2\pi I} \int_{-r_0}^x \dd x' \, f(x') \Pi (t+x'-x) \\
    & \simeq \tilde{J}_{a\bar{a}} (t,-r_0) + \frac{\delta_{a,\bar{a}}}{2\pi I} \Pi (t) (q(x)-q(-r_0)) \,, 
\end{aligned}
    \label{eq:approximation_tildeJ}
\end{align}
where we use a small core approximation from the first line to the second line (see Ref.~\cite{Polchinski:1984uw}).
By inserting Eq.~\eqref{eq:approximation_tildeJ} into Eq.~\eqref{eq:eom_for_Pi}, we obtain
\begin{align}
    \frac{\dd \Pi}{\dd t} \simeq \tilde{J} (t,-r_0) - \frac{N}{4\pi I} \Pi \,.
\end{align}
That is,
\begin{align}
    \Pi (t) = \mathrm{e}^{-\frac{N}{4\pi I}t} \Pi (0) + \int_0^t \dd t' \, \mathrm{e}^{-\frac{N}{4\pi I}(t-t')} \tilde{J} (t',-r_0) \,.
\end{align}
Assuming $\tilde{J}(t,-r_0)$ varies slowly in time and assuming $t \gg I$ so that the initial value of $\Pi (0)$ has decayed to essentially $0$,\footnote{This approximation is the same as the small core approximation.}
 we can approximate as 
\begin{align}
    \Pi (t) \simeq \frac{4\pi I}{N} \tilde{J} (t,-r_0) \,.
\label{eq:Pi}\end{align}
This result is consistent with the assumption we imposed.
By inserting Eq.~\eqref{eq:Pi} into Eq.~\eqref{eq:approximation_tildeJ} with setting $x=r_0$, we obtain
\begin{align}
    \left. \tilde{J}_{a\bar{a}} \right|_{x=r_0}  \simeq  \left. \tilde{J}_{a\bar{a}} \right|_{x=-r_0} - \frac{2}{N} \left. \tilde{J} \right|_{x=-r_0} \,.
        \label{eq:baryon_aa_bc}
\end{align}
This is the same as the dyon boundary condition \eqref{dyon_bc}.
We can also obtain a boundary condition for $\tilde{J}$ as 
\begin{align}
    \left. \tilde{J} \right|_{x=r_0} \simeq - \left. \tilde{J} \right|_{x=-r_0} \,.
\label{eq:baryon_bc}\end{align}
$\tilde{J}(t,x)$ changes the sign when it passes the dyon, and this implies a violation of fermion number conservation.

Finally, we derive a boundary condition for the $SU(N)$ current,
\begin{align}
    J_{SU(N)}^A := \chi_a^{\dagger} (T^A)_{a\bar{a}} \chi_{\bar{a}} \,.
\end{align}
In contrast to the $U(1)$ current, the $SU(N)$ current 
does not couple to the rotor and have no anomaly.
Thus, the boundary condition can be written as 
\begin{align}
    \left. J_{SU(N)}^A \right|_{x=r_0}  \simeq \left. J_{SU(N)}^A \right|_{x=-r_0}  \,.
    \label{eq:SU(n)current}
\end{align}

In this section, we derived boundary conditions imposed at the core for fermions \eqref{eq:bc_at_origin}, the (improved) $U(1)$ current \eqref{eq:baryon_bc} and the $SU(N)$ current \eqref{eq:SU(n)current}.

\subsection{Scattering process}

Let us consider a scattering process.
By integrating Eq.~\eqref{eq:baryon_aa_bc} over $t$, the boundary condition for each fermion number is derived:
\begin{align}
    n_a^\text{out}= n_a^\text{in}-\frac{2}{N}\sum_b n_b^\text{in} \,,
    \label{eq:change_fermion_number}
\end{align}
where $n_a^\text{out}$ and $n_a^\text{in}$ are the outgoing and incoming fermion number of $a$-th flavour, respectively,
\begin{align}
    n_a^\text{out} = \sum_{\bar{a}=1}^N  \int \dd t\, \tilde{J}_{a{\bar{a}}}(t,r_0) \delta_{a,\bar{a}} \,, \qquad 
    n_a^\text{in} = \sum_{\bar{a}=1}^N \int \dd t\, \tilde{J}_{a{\bar{a}}}(t,-r_0) \delta_{a,\bar{a}} \,.
\label{eq:n_def}\end{align}

This relation leads to the following conservation law in the monopole-fermions scattering for any number of $N$:
\begin{align}
    \sum_{a=1}^N n_a^{\text{out}} = -  \sum_{a=1}^N n_a^{\text{in}}\,.
\end{align}
Namely, the total number of the right-moving fermions 
is just flipped its sign via the scattering.
In the 4d language, it corresponds to the fermion number violating scattering.

\paragraph{$N=1$ case}
In this case, Eq.~\eqref{eq:change_fermion_number} reduces to\footnote{Here and hereafter we omit the subscript $a$ and $\bar{a}$ for $N=1$.}
\begin{align}
    n^\text{out}= - n^\text{in} \,.
\end{align}
For $n^\text{in} = 1$, we obtain $n^\text{out}=-1$.
This implies that the particle $\chi$ becomes the anti-particle $\chi^\dagger$ after the scattering.
That is, the incoming particle $\chi_\text{in}^-$ becomes the outgoing particle $\chi_\text{out}^-$.
The helicity of the incoming fermion changes after the scattering as in Ref.~\cite{Kazama:1976fm}. 
It is possible to replace the existence of the dyon with the boundary condition at the origin on the fermions and derive the same conclusion.
This is achieved by the boundary condition,
\begin{align}
    \left. \chi^\dagger \right|_{x = 0+} = \left. \mathrm{e}^{\i \varphi} \chi \right|_{x = 0-}  \,,
\end{align}
where $\varphi$ should be identified with a topological angle of the UV non-Abelian gauge theory.
This boundary condition is identical to Eq.~\eqref{eq:boundaryN1} under the complex conjugate.

\paragraph{$N=2$ case}

In this case, Eq.~\eqref{eq:change_fermion_number} reduces to 
\begin{align}
    n_a^\text{out}=  n_a^\text{in} - \sum_b n_b^\text{in} \,.
\end{align}
For $n_1^\text{in} = 1, n_2^\text{in} = 0$ case, we obtain $n_1^\text{out} = 0, n_2^\text{out} = -1$. 
The incoming particle $\chi_1$ becomes the outgoing anti-fermion $\chi_2^\dagger$.
Or equivalently, $\chi_{\text{in},1}^-$ becomes $\chi_{\text{out},2}^-$, and this implies that both the flavour and the chirality change after the scattering.

As in the $N = 1$ case, we can obtain the same result by imposing the boundary condition on the fermions,
\begin{align}
     \left. \epsilon_{\bar{a}b} \chi_b^{\dagger} \right|_{x = 0+} = \left. \i \, \mathrm{e}^{\i \varphi} \chi_{\bar{a}} \right|_{x = 0-} \,,
\end{align}
which is identical to Eq.~\eqref{eq:boundaryN2}  under the complex conjugate.

\paragraph{$N \geq 4$ cases}

Unlike the $N = 1,2$ cases, $n_a^\text{out}$ become fractional numbers in general when $n_a^\text{in}$ are integers.
This implies that the final state does not exist as an ordinary 2d fermion.
This is a unitarity paradox
in monopole-fermion scattering.
In Ref.~\cite{Polchinski:1984uw}, Polchinski concluded that the final state does not have a simple particle interpretation but appears as propagating pulses of vacuum polarisation.
In Sec.~\ref{sec:proposal}, we give our interpretation.

\paragraph{Change of $\alpha$}

From Eq.~\eqref{eq:Polchinski}, the equation for $\alpha$ is\footnote{From 4d perspective, Eq.~\eqref{Eq:alpha_change_derivative} may be derived from the anomaly equation, 
$\int \! \dd^3x\,\partial_\mu j^\mu \sim \int \!  \dd^3x \,\mathrm{Tr}\left(F\wedge F\right)$ \cite{Rubakov:1982fp}.}
\begin{align}
    \frac{1}{2\pi}\frac{\dd \alpha}{\dd t} = \frac{\Pi}{2\pi I}= \frac{2}{N} \tilde{J} (t,-r_0) \,,
    \label{Eq:alpha_change_derivative}
\end{align}
where Eqs.~\eqref{eq:Pi} and \eqref{eq:baryon_bc} are used in the last equality.
By integrating over $t$, we obtain
\begin{align}
    \frac{1}{2\pi}\left(\alpha_\text{out}-\alpha_\text{in}\right)= \frac{2}{N}\sum_a n_a^\text{in} \,,
\label{Eq:alpha_change}\end{align}
where $\alpha_\text{out}$ and $\alpha_\text{in}$ are the values of $\alpha$ after and before the scattering, respectively.
In the next section, we use the above relation.

\section{A four-dimensional solution to the monopole puzzle}
\label{sec:proposal}

In the previous section, we have seen that the state with the fractional fermion number indeed appears as the final state of the scattering.
The monopole puzzle is how to interpret this state in the 4d language.
We expect that, in the region far away from the monopole, the system is weakly interacting $U(1)$ gauge theory with the massless fermions. 
Then, the state with the fractional fermion number should belong to the Fock space of the fermions, but there are no fractional fermion states in the Fock space by construction.

Here, we propose a solution to the problem. We first point out that, if the rotor degrees of freedom, $\alpha$, is treated classically, then the problem is solved.
In this case, the fermion Fock space $H_\alpha$ is parameterised by $\alpha$. The initial fermion belongs to $H_{\alpha_\text{in}}$, while the final fermion belongs to the different Fock space $H_{\alpha_\text{out}}$. Here $\alpha_\text{in}$ and $\alpha_\text{out}$ are the values of $\alpha$ before and after the scattering, respectively.
In the following, in Sec.~\ref{sec:box}, we show that this picture nicely explains the final state of the monopole-fermion scattering for any $N=$ even with an IR cutoff.
Next, in Sec.~\ref{Sec:degenerate_vacua}, we turn to the question of the quantum effect on $\alpha$.
We argue that the puzzle is solved even if we take into account the quantum effect.

\subsection{The monopole scattering with the dynamics of \texorpdfstring{$\alpha$}{alpha}}
\label{sec:box}

We start from Eq.~\eqref{Eq:alpha_change}. 
The value of $\alpha$ after scattering is ($\alpha_\text{in}=0$ is assumed in this section)
\begin{align}
\alpha_\text{out} = \frac{4\pi}{N} \sum_a n_a^\text{in} \,.
\label{Eq:final_verphi}\end{align}
This simple formula provides insight into the monopole-fermion scattering.
When $N=1$ and $2$ cases,
$\alpha_\text{out} \equiv 0$ (mod $2\pi$) for any number of $\sum_a n_a^{\text{in}}$.
It is known that in the 2d phenomenological model 
there is the $2\pi$ periodicity of $\alpha$ 
up to the total fermion number, which would change by mod $2$ \cite{Polchinski:1984uw} (see also Footnote~\ref{footnote12}). 
On the other hand, 
the first non-trivial value appears  
$\alpha_\text{out} \equiv \pi$ (mod $2\pi$) when $N=4$ and $\sum_a n_a^{\text{in}}=\pm1$.
This is exactly the monopole-fermion scattering that suffers from the unitarity paradox.

Let us discuss the spectrum of the fermions.
We first recall 
Eq.~\eqref{eq:bc_at_origin}, which gives rise the phase shift at the monopole core.
In addition to that, we impose the IR cutoff $L$.
We place a boundary condition 
\begin{align}
    \left.\chi_{\bar{a}}\right|_{x=-L/2}=-\left.\chi_{\bar{a}} \right|_{x=L/2} \,,
\label{Eq:anti-periodic}\end{align}
so that
the momentum of the fermion \eqref{Eq:anti-periodic} for $\alpha_\text{in}=0$ is
\begin{align}
k_a = (2\ell_a+1)\frac{\pi}{L} \,, \qquad 
\ell_a\in \mathbb{Z} \,.
\label{Eq:initial_momentum}\end{align}
Following a Dirac sea picture, all states with $k_a<0$ (\emph{i.e.}, $\ell_a\leq-1$) are occupied in the vacuum. States with $k_a>0$ are considered to be particles while holes of $k_a<0$ states are understood as anti-particles.\footnote{We could think of this as placing an \emph{anti-monopole} at $x=L/2$ to cancel the tadpole of magnetic charge. We can set the anti-dyon degrees of freedom to $\beta=\pi$ before the fermion bounces back to $x=L/2$. We can also place an axial boundary condition. This is better as the system will be free of Majorana zero modes. This does not change our conclusions, though, except that we need to take $L\to 2L$.}

After the monopole-fermion scattering, $\alpha$ becomes Eq.~\eqref{Eq:final_verphi}.
Now the momentum compatible with Eqs.~\eqref{eq:bc_at_origin} and \eqref{Eq:anti-periodic} is
\begin{align}
k_a = \left(2\ell_a+1-\frac{4}{N} \sum_b n_b^\text{in} \right)\frac{\pi}{L} \,.
\label{Eq:after_momentum}\end{align}
In the rest of this section, we clarify that a candidate of the final state is obtained from the shift of the momentum.

\paragraph{$N=1$ and $2$ cases}
We observe that the whole spectra does not change for $N=1$ and $2$.
This implies that the monopole-fermion scattering is viewed as the spectral flow in these cases.
Suppose that $\ell\leq0$ states are initially occupied for $N = 1$.
That is, the $\ell = 0$ state corresponds to a particle, and the $\ell < 0$ states correspond to a Dirac sea.
After the scattering, by using Eq.~\eqref{Eq:after_momentum} with $n^\text{in}=1$, the state with $k=-\pi/L$ (corresponding to $\ell=1$) is not occupied, which is interpreted as an anti-particle.
Explicitly, the scattering process is written as
\begin{align}
    M + \chi_{\text{in}}^- 
    \to M + \chi_{\text{out}}^- \,.
\end{align}
Note that this needs not to be the true final state since the scattering process is not adiabatic. 
Indeed, the typical time scale of the scattering is set by the inverse of the GUT scale, which is much shorter than the inverse of the energy gap $\sim 1/L$.
It is possible that final state contains the fermion anti-fermion pairs.
When the final state is a massless single particle as given by the boundary condition \eqref{eq:boundaryN1}, the pair-production must lead to the on-shell photon exchange which is forbidden in s-wave scattering.
The figure is just an illustration to show that there is a candidate for the final state.
Similarly, if the $\ell\leq-2$ states are occupied initially (corresponding to $n^\text{in}=-1$), the final state corresponds to a particle state:
\begin{align}
    M + \chi_{\text{in}}^+ 
    \to M + \chi_{\text{out}}^+ \,.
\end{align}
This is the same as Eq.~\eqref{eq:N=1}.

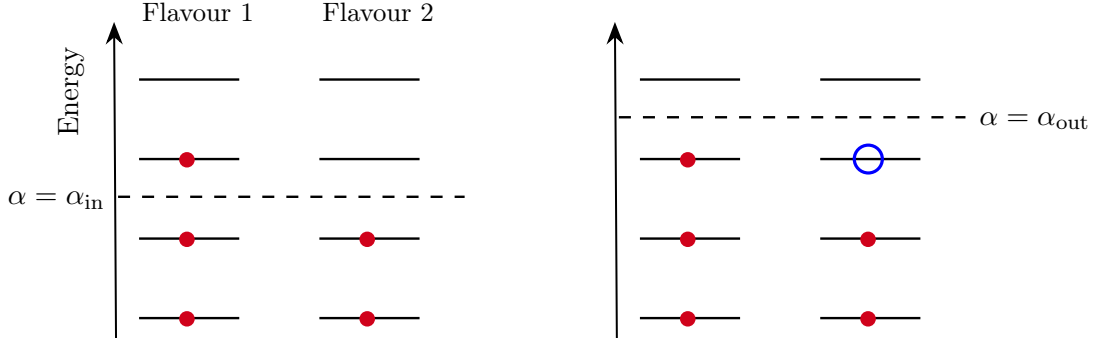
\begin{figure}[t!]
    \centering

\tikzset{every picture/.style={line width=0.75pt}} 

\begin{tikzpicture}[x=0.75pt,y=0.75pt,yscale=-1,xscale=1]

\draw [shift={(-135,-60)}, rotate =0] [draw opacity=1] [dash pattern={on 4.5pt off 4.5pt}] [line width=0.9]  (289.5,208.98) -- (462.5,208.98) ;
\draw [shift={(-135,-10)}, rotate =0] [draw opacity=1] [line width=0.9]  (300,220) -- (350,220) ;
\draw [shift={(-45,-10)}, rotate =0] [draw opacity=1] [line width=0.9]  (300,220) -- (350,220) ;
\draw [shift={(-135,-50)}, rotate =0] [draw opacity=1] [line width=0.9]  (300,220) -- (350,220) ;
\draw [shift={(-45,-50)}, rotate =0] [draw opacity=1] [line width=0.9]  (300,220) -- (350,220) ;
\draw [shift={(-135,-90)}, rotate =0] [draw opacity=1] [line width=0.9]  (300,220) -- (350,220) ;
\draw [shift={(-45,-90)}, rotate =0] [draw opacity=1] [line width=0.9]  (300,220) -- (350,220) ;
\draw [shift={(-135,-130)}, rotate =0] [draw opacity=1] [line width=0.9]  (300,220) -- (350,220) ;
\draw [shift={(-45,-130)}, rotate =0] [draw opacity=1] [line width=0.9]  (300,220) -- (350,220) ;

\draw  [shift={(-308,0)}, rotate =0] [color={rgb, 255:red, 208; green, 2; blue, 27 }  ,draw opacity=1 ][fill={rgb, 255:red, 208; green, 2; blue, 27 }  ,fill opacity=1 ] (493.5,170.38) .. controls (493.5,172.24) and (495.01,173.75) .. (496.88,173.75) .. controls (498.74,173.75) and (500.25,172.24) .. (500.25,170.38) .. controls (500.25,168.51) and (498.74,167) .. (496.88,167) .. controls (495.01,167) and (493.5,168.51) .. (493.5,170.38) -- cycle ;
\draw  [shift={(-308,40)}, rotate =0] [color={rgb, 255:red, 208; green, 2; blue, 27 }  ,draw opacity=1 ][fill={rgb, 255:red, 208; green, 2; blue, 27 }  ,fill opacity=1 ] (493.5,170.38) .. controls (493.5,172.24) and (495.01,173.75) .. (496.88,173.75) .. controls (498.74,173.75) and (500.25,172.24) .. (500.25,170.38) .. controls (500.25,168.51) and (498.74,167) .. (496.88,167) .. controls (495.01,167) and (493.5,168.51) .. (493.5,170.38) -- cycle ;
\draw  [shift={(-308,-40)}, rotate =0] [color={rgb, 255:red, 208; green, 2; blue, 27 }  ,draw opacity=1 ][fill={rgb, 255:red, 208; green, 2; blue, 27 }  ,fill opacity=1 ] (493.5,170.38) .. controls (493.5,172.24) and (495.01,173.75) .. (496.88,173.75) .. controls (498.74,173.75) and (500.25,172.24) .. (500.25,170.38) .. controls (500.25,168.51) and (498.74,167) .. (496.88,167) .. controls (495.01,167) and (493.5,168.51) .. (493.5,170.38) -- cycle ;
\draw  [shift={(-218,0)}, rotate =0] [color={rgb, 255:red, 208; green, 2; blue, 27 }  ,draw opacity=1 ][fill={rgb, 255:red, 208; green, 2; blue, 27 }  ,fill opacity=1 ] (493.5,170.38) .. controls (493.5,172.24) and (495.01,173.75) .. (496.88,173.75) .. controls (498.74,173.75) and (500.25,172.24) .. (500.25,170.38) .. controls (500.25,168.51) and (498.74,167) .. (496.88,167) .. controls (495.01,167) and (493.5,168.51) .. (493.5,170.38) -- cycle ;
\draw  [shift={(-218,40)}, rotate =0] [color={rgb, 255:red, 208; green, 2; blue, 27 }  ,draw opacity=1 ][fill={rgb, 255:red, 208; green, 2; blue, 27 }  ,fill opacity=1 ] (493.5,170.38) .. controls (493.5,172.24) and (495.01,173.75) .. (496.88,173.75) .. controls (498.74,173.75) and (500.25,172.24) .. (500.25,170.38) .. controls (500.25,168.51) and (498.74,167) .. (496.88,167) .. controls (495.01,167) and (493.5,168.51) .. (493.5,170.38) -- cycle ;

\draw [-{Stealth[length=3mm]}] (153.48,220) -- (152.52,61) ;

\draw (123,120) node [anchor=north west][inner sep=0.75pt]  [rotate=-270] [align=left] {{Energy}};
\draw (98,145) node [anchor=north west][inner sep=0.75pt]  [rotate=0] [align=left] {{$\alpha=\alpha_\text{in}$}};
\draw (583,105) node [anchor=north west][inner sep=0.75pt]  [rotate=0] [align=left] {{$\alpha=\alpha_\text{out}$}};

\draw (165,50) node [anchor=north west][inner sep=0.75pt]  [opacity=1 ] [align=left] {\textcolor[rgb]{0,0,0}{{\fontsize{10pt}{10pt}\selectfont Flavour 1}}};
\draw (255,50) node [anchor=north west][inner sep=0.75pt]  [opacity=1 ] [align=left] {\textcolor[rgb]{0,0,0}{{\fontsize{10pt}{10pt}\selectfont Flavour 2}}};

\draw [shift={(-135+250,-60-40)}, rotate =0] [draw opacity=1] [dash pattern={on 4.5pt off 4.5pt}] [line width=0.9]  (289.5,208.98) -- (462.5,208.98) ;
\draw [shift={(-135+250,-10)}, rotate =0] [draw opacity=1] [line width=0.9]  (300,220) -- (350,220) ;
\draw [shift={(-45+250,-10)}, rotate =0] [draw opacity=1] [line width=0.9]  (300,220) -- (350,220) ;
\draw [shift={(-135+250,-50)}, rotate =0] [draw opacity=1] [line width=0.9]  (300,220) -- (350,220) ;
\draw [shift={(-45+250,-50)}, rotate =0] [draw opacity=1] [line width=0.9]  (300,220) -- (350,220) ;
\draw [shift={(-135+250,-90)}, rotate =0] [draw opacity=1] [line width=0.9]  (300,220) -- (350,220) ;
\draw [shift={(-45+250,-90)}, rotate =0] [draw opacity=1] [line width=0.9]  (300,220) -- (350,220) ;
\draw [shift={(-135+250,-130)}, rotate =0] [draw opacity=1] [line width=0.9]  (300,220) -- (350,220) ;
\draw [shift={(-45+250,-130)}, rotate =0] [draw opacity=1] [line width=0.9]  (300,220) -- (350,220) ;

\draw [-{Stealth[length=3mm]},shift={(250,0)}, rotate = 0] (153.48,220) -- (152.52,61) ;

\draw  [shift={(-308+250,0)}, rotate =0] [color={rgb, 255:red, 208; green, 2; blue, 27 }  ,draw opacity=1 ][fill={rgb, 255:red, 208; green, 2; blue, 27 }  ,fill opacity=1 ] (493.5,170.38) .. controls (493.5,172.24) and (495.01,173.75) .. (496.88,173.75) .. controls (498.74,173.75) and (500.25,172.24) .. (500.25,170.38) .. controls (500.25,168.51) and (498.74,167) .. (496.88,167) .. controls (495.01,167) and (493.5,168.51) .. (493.5,170.38) -- cycle ;
\draw  [shift={(-308+250,40)}, rotate =0] [color={rgb, 255:red, 208; green, 2; blue, 27 }  ,draw opacity=1 ][fill={rgb, 255:red, 208; green, 2; blue, 27 }  ,fill opacity=1 ] (493.5,170.38) .. controls (493.5,172.24) and (495.01,173.75) .. (496.88,173.75) .. controls (498.74,173.75) and (500.25,172.24) .. (500.25,170.38) .. controls (500.25,168.51) and (498.74,167) .. (496.88,167) .. controls (495.01,167) and (493.5,168.51) .. (493.5,170.38) -- cycle ;
\draw  [shift={(-218+250,40)}, rotate =0] [color={rgb, 255:red, 208; green, 2; blue, 27 }  ,draw opacity=1 ][fill={rgb, 255:red, 208; green, 2; blue, 27 }  ,fill opacity=1 ] (493.5,170.38) .. controls (493.5,172.24) and (495.01,173.75) .. (496.88,173.75) .. controls (498.74,173.75) and (500.25,172.24) .. (500.25,170.38) .. controls (500.25,168.51) and (498.74,167) .. (496.88,167) .. controls (495.01,167) and (493.5,168.51) .. (493.5,170.38) -- cycle ;
\draw  [shift={(-308+250,-40)}, rotate =0] [color={rgb, 255:red, 208; green, 2; blue, 27 }  ,draw opacity=1 ][fill={rgb, 255:red, 208; green, 2; blue, 27 }  ,fill opacity=1 ] (493.5,170.38) .. controls (493.5,172.24) and (495.01,173.75) .. (496.88,173.75) .. controls (498.74,173.75) and (500.25,172.24) .. (500.25,170.38) .. controls (500.25,168.51) and (498.74,167) .. (496.88,167) .. controls (495.01,167) and (493.5,168.51) .. (493.5,170.38) -- cycle ;
\draw  [shift={(-218+250,0)}, rotate =0] [color={rgb, 255:red, 208; green, 2; blue, 27 }  ,draw opacity=1 ][fill={rgb, 255:red, 208; green, 2; blue, 27 } ,fill opacity=1 ] (493.5,170.38) .. controls (493.5,172.24) and (495.01,173.75) .. (496.88,173.75) .. controls (498.74,173.75) and (500.25,172.24) .. (500.25,170.38) .. controls (500.25,168.51) and (498.74,167) .. (496.88,167) .. controls (495.01,167) and (493.5,168.51) .. (493.5,170.38) -- cycle ;
\draw [color=blue, very thick] (529,130) circle [radius=7.0];
\end{tikzpicture}

    \caption{The fermion spectrum before and after the scattering for $N=2$. Left: Initial state; $n_1^\text{in}=1, n_2^\text{in}=0$. All the negative states and the one flavour 1 positive energy state are occupied. Right: A candidate of the final state; $n_1^\text{out}=0, n_2^\text{out}=-1$, which is interpreted as one anti-particle state of flavour 2 and is represented by the blue circle. We emphasise that, since the scattering is not adiabatic, this is not necessarily a true final state. The point here is that there are candidates for the final state.}
    \label{Fig:anomaly_Nf2}
\end{figure}

For $N=2$, when an initial state is $\chi_{\text{in},\bar{1}}^-$ ($\ell_1\leq0$ and $\ell_2\leq-1$ states are occupied), the final state is $\chi_{\text{out},2}^-$ following Eq.~\eqref{Eq:after_momentum} with $n_1^\text{in}=1, n_2^\text{in}=0$:
\begin{align}
    M +\chi_{\text{in},\bar{1}}^-\to M +\chi_{\text{out},2}^- \,,
\end{align}
see also Fig.~\ref{Fig:anomaly_Nf2}.
The result same as Eq.~\eqref{eq:N=2} is obtained by choosing an initial state where $\ell_1\leq-2$ and $\ell_2\leq-1$ are occupied (corresponding to $n^\text{in}_1=-1, n^\text{in}_2=0$).

In other words, for $N=1$ and $2$ cases,
the fermion momentum after the monopole-fermion scattering is still the same as the anti-periodic boundary condition \eqref{Eq:after_momentum}. 
Hence, the fermion Fock space after the scattering does not change from the initial one.

\begin{figure}[t]
    \centering

\tikzset{every picture/.style={line width=0.75pt}} 

\begin{tikzpicture}[x=0.75pt,y=0.75pt,yscale=-1,xscale=1]

\draw [shift={(-265,-60)}, rotate =0] [draw opacity=1] [dash pattern={on 4.5pt off 4.5pt}] [line width=0.9]  (289.5,208.98) -- (289.5+65*4,208.98) ;
\draw [shift={(-260,-10)}, rotate =0] [draw opacity=1] [line width=0.9]  (300,220) -- (340,220) ;
\draw [shift={(-260,-50)}, rotate =0] [draw opacity=1] [line width=0.9]  (300,220) -- (340,220) ;
\draw [shift={(-260,-90)}, rotate =0] [draw opacity=1] [line width=0.9]  (300,220) -- (340,220) ;
\draw [shift={(-260,-130)}, rotate =0] [draw opacity=1] [line width=0.9]  (300,220) -- (340,220) ;
\draw [shift={(-260+65,-10)}, rotate =0] [draw opacity=1] [line width=0.9]  (300,220) -- (340,220) ;
\draw [shift={(-260+65,-50)}, rotate =0] [draw opacity=1] [line width=0.9]  (300,220) -- (340,220) ;
\draw [shift={(-260+65,-90)}, rotate =0] [draw opacity=1] [line width=0.9]  (300,220) -- (340,220) ;
\draw [shift={(-260+65,-130)}, rotate =0] [draw opacity=1] [line width=0.9]  (300,220) -- (340,220) ;
\draw [shift={(-260+65*2,-10)}, rotate =0] [draw opacity=1] [line width=0.9]  (300,220) -- (340,220) ;
\draw [shift={(-260+65*2,-50)}, rotate =0] [draw opacity=1] [line width=0.9]  (300,220) -- (340,220) ;
\draw [shift={(-260+65*2,-90)}, rotate =0] [draw opacity=1] [line width=0.9]  (300,220) -- (340,220) ;
\draw [shift={(-260+65*2,-130)}, rotate =0] [draw opacity=1] [line width=0.9]  (300,220) -- (340,220) ;
\draw [shift={(-260+65*3,-10)}, rotate =0] [draw opacity=1] [line width=0.9]  (300,220) -- (340,220) ;
\draw [shift={(-260+65*3,-50)}, rotate =0] [draw opacity=1] [line width=0.9]  (300,220) -- (340,220) ;
\draw [shift={(-260+65*3,-90)}, rotate =0] [draw opacity=1] [line width=0.9]  (300,220) -- (340,220) ;
\draw [shift={(-260+65*3,-130)}, rotate =0] [draw opacity=1] [line width=0.9]  (300,220) -- (340,220) ;

\draw  [shift={(-438,0)}, rotate =0] [color={rgb, 255:red, 208; green, 2; blue, 27 }  ,draw opacity=1 ][fill={rgb, 255:red, 208; green, 2; blue, 27 }  ,fill opacity=1 ] (493.5,170.38) .. controls (493.5,172.24) and (495.01,173.75) .. (496.88,173.75) .. controls (498.74,173.75) and (500.25,172.24) .. (500.25,170.38) .. controls (500.25,168.51) and (498.74,167) .. (496.88,167) .. controls (495.01,167) and (493.5,168.51) .. (493.5,170.38) -- cycle ;
\draw  [shift={(-438,40)}, rotate =0] [color={rgb, 255:red, 208; green, 2; blue, 27 }  ,draw opacity=1 ][fill={rgb, 255:red, 208; green, 2; blue, 27 }  ,fill opacity=1 ] (493.5,170.38) .. controls (493.5,172.24) and (495.01,173.75) .. (496.88,173.75) .. controls (498.74,173.75) and (500.25,172.24) .. (500.25,170.38) .. controls (500.25,168.51) and (498.74,167) .. (496.88,167) .. controls (495.01,167) and (493.5,168.51) .. (493.5,170.38) -- cycle ;
\draw  [shift={(-438,-40)}, rotate =0] [color={rgb, 255:red, 208; green, 2; blue, 27 }  ,draw opacity=1 ][fill={rgb, 255:red, 208; green, 2; blue, 27 }  ,fill opacity=1 ] (493.5,170.38) .. controls (493.5,172.24) and (495.01,173.75) .. (496.88,173.75) .. controls (498.74,173.75) and (500.25,172.24) .. (500.25,170.38) .. controls (500.25,168.51) and (498.74,167) .. (496.88,167) .. controls (495.01,167) and (493.5,168.51) .. (493.5,170.38) -- cycle ;
\draw  [shift={(-438+65,0)}, rotate =0] [color={rgb, 255:red, 208; green, 2; blue, 27 }  ,draw opacity=1 ][fill={rgb, 255:red, 208; green, 2; blue, 27 }  ,fill opacity=1 ] (493.5,170.38) .. controls (493.5,172.24) and (495.01,173.75) .. (496.88,173.75) .. controls (498.74,173.75) and (500.25,172.24) .. (500.25,170.38) .. controls (500.25,168.51) and (498.74,167) .. (496.88,167) .. controls (495.01,167) and (493.5,168.51) .. (493.5,170.38) -- cycle ;
\draw  [shift={(-438+65,40)}, rotate =0] [color={rgb, 255:red, 208; green, 2; blue, 27 }  ,draw opacity=1 ][fill={rgb, 255:red, 208; green, 2; blue, 27 }  ,fill opacity=1 ] (493.5,170.38) .. controls (493.5,172.24) and (495.01,173.75) .. (496.88,173.75) .. controls (498.74,173.75) and (500.25,172.24) .. (500.25,170.38) .. controls (500.25,168.51) and (498.74,167) .. (496.88,167) .. controls (495.01,167) and (493.5,168.51) .. (493.5,170.38) -- cycle ;
\draw  [shift={(-438+65*2,0)}, rotate =0] [color={rgb, 255:red, 208; green, 2; blue, 27 }  ,draw opacity=1 ][fill={rgb, 255:red, 208; green, 2; blue, 27 }  ,fill opacity=1 ] (493.5,170.38) .. controls (493.5,172.24) and (495.01,173.75) .. (496.88,173.75) .. controls (498.74,173.75) and (500.25,172.24) .. (500.25,170.38) .. controls (500.25,168.51) and (498.74,167) .. (496.88,167) .. controls (495.01,167) and (493.5,168.51) .. (493.5,170.38) -- cycle ;
\draw  [shift={(-438+65*2,40)}, rotate =0] [color={rgb, 255:red, 208; green, 2; blue, 27 }  ,draw opacity=1 ][fill={rgb, 255:red, 208; green, 2; blue, 27 }  ,fill opacity=1 ] (493.5,170.38) .. controls (493.5,172.24) and (495.01,173.75) .. (496.88,173.75) .. controls (498.74,173.75) and (500.25,172.24) .. (500.25,170.38) .. controls (500.25,168.51) and (498.74,167) .. (496.88,167) .. controls (495.01,167) and (493.5,168.51) .. (493.5,170.38) -- cycle ;
\draw  [shift={(-438+65*3,0)}, rotate =0] [color={rgb, 255:red, 208; green, 2; blue, 27 }  ,draw opacity=1 ][fill={rgb, 255:red, 208; green, 2; blue, 27 }  ,fill opacity=1 ] (493.5,170.38) .. controls (493.5,172.24) and (495.01,173.75) .. (496.88,173.75) .. controls (498.74,173.75) and (500.25,172.24) .. (500.25,170.38) .. controls (500.25,168.51) and (498.74,167) .. (496.88,167) .. controls (495.01,167) and (493.5,168.51) .. (493.5,170.38) -- cycle ;
\draw  [shift={(-438+65*3,40)}, rotate =0] [color={rgb, 255:red, 208; green, 2; blue, 27 }  ,draw opacity=1 ][fill={rgb, 255:red, 208; green, 2; blue, 27 }  ,fill opacity=1 ] (493.5,170.38) .. controls (493.5,172.24) and (495.01,173.75) .. (496.88,173.75) .. controls (498.74,173.75) and (500.25,172.24) .. (500.25,170.38) .. controls (500.25,168.51) and (498.74,167) .. (496.88,167) .. controls (495.01,167) and (493.5,168.51) .. (493.5,170.38) -- cycle ;

\draw [-{Stealth[length=3mm]}, shift={(-130,0)}, rotate = 0] [draw opacity=1 ]   (153.48,220) -- (152.52,61) ;

\draw [shift={(-125,0)}, rotate = 0] (120.5,151.5) node [anchor=north west][inner sep=0.75pt]  [rotate=-270] [align=left] {{Energy}};

\draw (37,50) node [anchor=north west][inner sep=0.75pt]  [opacity=1 ] [align=left] {\textcolor[rgb]{0,0,0}{{\fontsize{9pt}{9pt}\selectfont Flavour 1}}};
\draw (37+65,50) node [anchor=north west][inner sep=0.75pt]  [opacity=1 ] [align=left] {\textcolor[rgb]{0,0,0}{{\fontsize{9pt}{9pt}\selectfont Flavour 2}}};
\draw (37+65*2,50) node [anchor=north west][inner sep=0.75pt]  [opacity=1 ] [align=left] {\textcolor[rgb]{0,0,0}{{\fontsize{9pt}{9pt}\selectfont Flavour 3}}};
\draw (37+65*3,50) node [anchor=north west][inner sep=0.75pt]  [opacity=1 ] [align=left] {\textcolor[rgb]{0,0,0}{{\fontsize{9pt}{9pt}\selectfont Flavour 4}}};

\draw [shift={(-265+300,-80)}, rotate =0] [draw opacity=1] [dash pattern={on 4.5pt off 4.5pt}] [line width=0.9]  (289.5,208.98) -- (289.5+65*4,208.98) ;
\draw [shift={(-260+300,-10)}, rotate =0] [draw opacity=1] [line width=0.9]  (300,220) -- (340,220) ;
\draw [shift={(-260+300,-50)}, rotate =0] [draw opacity=1] [line width=0.9]  (300,220) -- (340,220) ;
\draw [shift={(-260+300,-90)}, rotate =0] [draw opacity=1] [line width=0.9]  (300,220) -- (340,220) ;
\draw [shift={(-260+300,-130)}, rotate =0] [draw opacity=1] [line width=0.9]  (300,220) -- (340,220) ;
\draw [shift={(-260+65+300,-10)}, rotate =0] [draw opacity=1] [line width=0.9]  (300,220) -- (340,220) ;
\draw [shift={(-260+65+300,-50)}, rotate =0] [draw opacity=1] [line width=0.9]  (300,220) -- (340,220) ;
\draw [shift={(-260+65+300,-90)}, rotate =0] [draw opacity=1] [line width=0.9]  (300,220) -- (340,220) ;
\draw [shift={(-260+65+300,-130)}, rotate =0] [draw opacity=1] [line width=0.9]  (300,220) -- (340,220) ;
\draw [shift={(-260+65*2+300,-10)}, rotate =0] [draw opacity=1] [line width=0.9]  (300,220) -- (340,220) ;
\draw [shift={(-260+65*2+300,-50)}, rotate =0] [draw opacity=1] [line width=0.9]  (300,220) -- (340,220) ;
\draw [shift={(-260+65*2+300,-90)}, rotate =0] [draw opacity=1] [line width=0.9]  (300,220) -- (340,220) ;
\draw [shift={(-260+65*2+300,-130)}, rotate =0] [draw opacity=1] [line width=0.9]  (300,220) -- (340,220) ;
\draw [shift={(-260+65*3+300,-10)}, rotate =0] [draw opacity=1] [line width=0.9]  (300,220) -- (340,220) ;
\draw [shift={(-260+65*3+300,-50)}, rotate =0] [draw opacity=1] [line width=0.9]  (300,220) -- (340,220) ;
\draw [shift={(-260+65*3+300,-90)}, rotate =0] [draw opacity=1] [line width=0.9]  (300,220) -- (340,220) ;
\draw [shift={(-260+65*3+300,-130)}, rotate =0] [draw opacity=1] [line width=0.9]  (300,220) -- (340,220) ;
\draw  [shift={(-438+300,-40)}, rotate =0] [color={rgb, 255:red, 208; green, 2; blue, 27 }  ,draw opacity=1 ][fill={rgb, 255:red, 208; green, 2; blue, 27 }  ,fill opacity=1 ] (493.5,170.38) .. controls (493.5,172.24) and (495.01,173.75) .. (496.88,173.75) .. controls (498.74,173.75) and (500.25,172.24) .. (500.25,170.38) .. controls (500.25,168.51) and (498.74,167) .. (496.88,167) .. controls (495.01,167) and (493.5,168.51) .. (493.5,170.38) -- cycle ;
\draw  [shift={(-438+300,0)}, rotate =0] [color={rgb, 255:red, 208; green, 2; blue, 27 }  ,draw opacity=1 ][fill={rgb, 255:red, 208; green, 2; blue, 27 }  ,fill opacity=1 ] (493.5,170.38) .. controls (493.5,172.24) and (495.01,173.75) .. (496.88,173.75) .. controls (498.74,173.75) and (500.25,172.24) .. (500.25,170.38) .. controls (500.25,168.51) and (498.74,167) .. (496.88,167) .. controls (495.01,167) and (493.5,168.51) .. (493.5,170.38) -- cycle ;
\draw  [shift={(-438+300,40)}, rotate =0] [color={rgb, 255:red, 208; green, 2; blue, 27 }  ,draw opacity=1 ][fill={rgb, 255:red, 208; green, 2; blue, 27 }  ,fill opacity=1 ] (493.5,170.38) .. controls (493.5,172.24) and (495.01,173.75) .. (496.88,173.75) .. controls (498.74,173.75) and (500.25,172.24) .. (500.25,170.38) .. controls (500.25,168.51) and (498.74,167) .. (496.88,167) .. controls (495.01,167) and (493.5,168.51) .. (493.5,170.38) -- cycle ;
\draw  [shift={(-438+65+300,0)}, rotate =0] [color={rgb, 255:red, 208; green, 2; blue, 27 }  ,draw opacity=1 ][fill={rgb, 255:red, 208; green, 2; blue, 27 }  ,fill opacity=1 ] (493.5,170.38) .. controls (493.5,172.24) and (495.01,173.75) .. (496.88,173.75) .. controls (498.74,173.75) and (500.25,172.24) .. (500.25,170.38) .. controls (500.25,168.51) and (498.74,167) .. (496.88,167) .. controls (495.01,167) and (493.5,168.51) .. (493.5,170.38) -- cycle ;
\draw  [shift={(-438+65+300,40)}, rotate =0] [color={rgb, 255:red, 208; green, 2; blue, 27 }  ,draw opacity=1 ][fill={rgb, 255:red, 208; green, 2; blue, 27 }  ,fill opacity=1 ] (493.5,170.38) .. controls (493.5,172.24) and (495.01,173.75) .. (496.88,173.75) .. controls (498.74,173.75) and (500.25,172.24) .. (500.25,170.38) .. controls (500.25,168.51) and (498.74,167) .. (496.88,167) .. controls (495.01,167) and (493.5,168.51) .. (493.5,170.38) -- cycle ;
\draw  [shift={(-438+65*2+300,0)}, rotate =0] [color={rgb, 255:red, 208; green, 2; blue, 27 }  ,draw opacity=1 ][fill={rgb, 255:red, 208; green, 2; blue, 27 }  ,fill opacity=1 ] (493.5,170.38) .. controls (493.5,172.24) and (495.01,173.75) .. (496.88,173.75) .. controls (498.74,173.75) and (500.25,172.24) .. (500.25,170.38) .. controls (500.25,168.51) and (498.74,167) .. (496.88,167) .. controls (495.01,167) and (493.5,168.51) .. (493.5,170.38) -- cycle ;
\draw  [shift={(-438+65*2+300,40)}, rotate =0] [color={rgb, 255:red, 208; green, 2; blue, 27 }  ,draw opacity=1 ][fill={rgb, 255:red, 208; green, 2; blue, 27 }  ,fill opacity=1 ] (493.5,170.38) .. controls (493.5,172.24) and (495.01,173.75) .. (496.88,173.75) .. controls (498.74,173.75) and (500.25,172.24) .. (500.25,170.38) .. controls (500.25,168.51) and (498.74,167) .. (496.88,167) .. controls (495.01,167) and (493.5,168.51) .. (493.5,170.38) -- cycle ;
\draw  [shift={(-438+65*3+300,0)}, rotate =0] [color={rgb, 255:red, 208; green, 2; blue, 27 }  ,draw opacity=1 ][fill={rgb, 255:red, 208; green, 2; blue, 27 }  ,fill opacity=1 ] (493.5,170.38) .. controls (493.5,172.24) and (495.01,173.75) .. (496.88,173.75) .. controls (498.74,173.75) and (500.25,172.24) .. (500.25,170.38) .. controls (500.25,168.51) and (498.74,167) .. (496.88,167) .. controls (495.01,167) and (493.5,168.51) .. (493.5,170.38) -- cycle ;
\draw  [shift={(-438+65*3+300,40)}, rotate =0] [color={rgb, 255:red, 208; green, 2; blue, 27 }  ,draw opacity=1 ][fill={rgb, 255:red, 208; green, 2; blue, 27 }  ,fill opacity=1 ] (493.5,170.38) .. controls (493.5,172.24) and (495.01,173.75) .. (496.88,173.75) .. controls (498.74,173.75) and (500.25,172.24) .. (500.25,170.38) .. controls (500.25,168.51) and (498.74,167) .. (496.88,167) .. controls (495.01,167) and (493.5,168.51) .. (493.5,170.38) -- cycle ;
\draw [-{Stealth[length=3mm]}, shift={(-130+300,0)}, rotate = 0] [draw opacity=1 ]   (153.48,220) -- (152.52,61) ;

\end{tikzpicture}
    \caption{The fermion spectrum before and after the scattering for $N=4$. Left: Initial state;  $n_1^\text{in}=1$, $n_{2,3,4}^\text{in}=0$. Right: A candidate of the final state; 
    $n_1^\text{out}=1/2$, $n_{2,3,4}^\text{out}=-1/2$, where 
    the Fock vacuum after the scattering has $\pm1/2$ fermion number as the vacuum contribution. This is interpreted as the semiton state.}
    \label{Fig:anomaly_Nf4}
\end{figure}
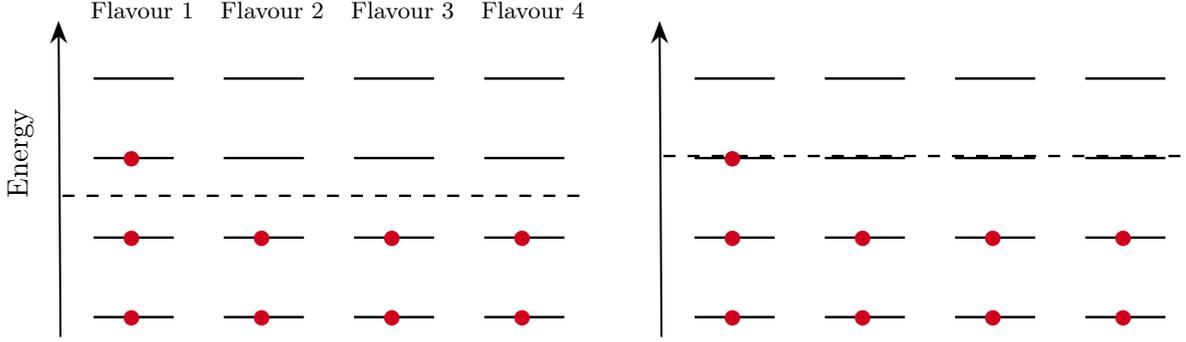

\paragraph{$N=4$ case}
Contrary to the $N=1$ and $2$ cases, the situation is drastically different for $N=4$. 
The whole spectra \eqref{Eq:initial_momentum} and \eqref{Eq:after_momentum} are different.
This means that the fermion state after the scattering can not be described by the initial fermion Fock space. 
From Eq.~\eqref{Eq:after_momentum}, we can see that the allowed momentum for $N=4$ is the same as  
the one of periodic (anti-periodic) boundary condition 
when $\sum_b n_b^{\text{in}}$ is an odd (even) number. Namely, 
the fermion one-particle state with anti-periodic condition is scattered into fermion one-particle state with periodic condition, see Fig.~\ref{Fig:anomaly_Nf4}.
In fact, this is related to the observation in Ref.~\cite{Maldacena:1995pq}, where it was shown that there are no unitarity paradox if one prepares both of the periodic and anti-periodic fermions from the beginning.
On the other hand, we solve the problem even if there is only a fermion with anti-periodic boundary condition, thanks to the dynamics of the rotor degree of freedom. 
This final state has the correct quantum number same as the semiton~\cite{Maldacena:1995pq} (see also Sec.~\ref{Sec:fermion_number}). 
We have provided a simple 4d picture, where $\alpha$ is just a modulus of the monopole.
Even though the model itself is 2d, the uplifting to 4d degrees of freedom is clear.

\paragraph{$N\geq6$ cases} As in the $N=4$ case, the whole spectra \eqref{Eq:initial_momentum} and \eqref{Eq:after_momentum} are different.
For general $N$, it is expected that the final state is described using the fermion $\zeta$ with the twisted boundary condition
\begin{align}
\zeta \sim - \mathrm{e}^{-2\pi \i \frac{2}{N}}\zeta \,,  
\label{Eq:twisted}\end{align}
for $n_{\text{in},a}=\delta_{a,1}$.
From the state-operator correspondence, the Fock vacuum of the twisted boundary condition \eqref{Eq:twisted} is expressed by the operator $\mathrm{e}^{-\i \frac{2}{N} \rho}$ (\eg, Ref.~\cite{Polchinski:1998rr}, see also Sec.~\ref{Sec:fermion_number}), where $\rho$ is the bosonised field of $\zeta$.
This indicates that the Fock vacuum before scattering has zero fermion number while the vacuum after scattering has $-2/N$ fermion number.
This provides an explanation of 
Eq.~\eqref{eq:change_fermion_number}. The first term in the right hand side is ``particle contribution'', and the second term is ``vacuum contribution''. The vacuum contribution can be fractional. 

To summarise, the unitarity paradox
is solved by treating $\alpha$ classically.
In our solution, the outgoing state is a state which belongs to a different Fock space.

\subsection{Fermion number}\label{Sec:fermion_number}
Here we clarify why the fermions with different boundary condition have the fractional fermion number.
First, we define 
\begin{align}
    \tilde{\chi}_{\bar{a}}=  
    \begin{dcases}
   \chi_{\bar{a}} \,, & (x >0)\,, \\
    \mathrm{e}^{ \i \alpha} \chi_{\bar{a}}\,, & (x <0)\,,
    \end{dcases}
\end{align}
such that $\tilde{\chi}_{\bar{a}}|_{x=-0} = \tilde{\chi}_{\bar{a}}|_{x=+0}$, 
which is consistent with  Eqs.~\eqref{eq:unfolding}  and \eqref{eq:bc_at_origin}.
Combined with Eq.~\eqref{Eq:anti-periodic}, $\tilde{\chi}$ satisfies
\begin{align}
    \left.\tilde{\chi}_{\bar{a}}\right|_{x=-L/2}=-\mathrm{e}^{\i\alpha}\left.\tilde{\chi}_{\bar{a}} \right|_{x=L/2} \,.
\label{Eq:bc_chi_tilde}\end{align}
By moving to the left-moving basis $\psi_{\mathrm{L}a}$, we obtain
\begin{align}
    \left.\psi_{\mathrm{L}a}\right|_{x=-L/2}=-\mathrm{e}^{- \i \alpha}\left.\psi_{\mathrm{L}a} \right|_{x=L/2} \,.
\label{Eq:Left_bc}\end{align}
The bosonisation is (see, \eg,~Ref.~\cite{Green:1987sp})
\begin{align}
\begin{aligned}
    \psi_{\mathrm{L}a}& = \, :\mathrm{e}^{\i \phi_{\mathrm{L}a}}: \\
    &= \exp \left(\sum_{n>0,n\in\mathbb{Z}} \frac{\mathrm{e}^{\i n\sigma}}{n}{\phi_{-n,a}}\right)\mathrm{e}^{\i{\phi_{0,a}}}\mathrm{e}^{\i\sigma\left({p_{0,a}}+\frac{1}{2}\right)} \exp \left(-\sum_{n>0,n\in\mathbb{Z}}\frac{\mathrm{e}^{-\i n\sigma}}{n} \phi_{n,a} \right)\,,
\end{aligned}
\end{align}
where $\sigma=2\pi x/L$, and $\phi_{\mathrm{L}}$ is the chiral boson:
 \begin{align}
     \phi_{\mathrm{L}a} = \phi_{0,a} +(\sigma+\tau){p_{0,a}} + \sum_{n\neq0,n\in \mathbb{Z}}{\phi_{n,a}} \mathrm{e}^{-\i n(\sigma+\tau)} \,.
 \end{align}
Here $\tau$ is the time coordinate, and the commutation relation is $[{p_{0,a}},{\phi_{0,b}}]=-\i \delta_{ab}, [\phi_n,\phi_m]=n\delta_{n,-m}$.
The momentum ${p_{0,a}}$ is identified as the $a$-th fermion number operator. 
Given the boundary condition \eqref{Eq:Left_bc}, this is quantised as
\begin{align}
\label{Eq:fermionnumber}
    \text{($a$-th fermion number)}=\frac{\alpha}{2\pi}+\mathbb{Z}\,.
\end{align}
This leads to the fractionalisation of the fermion number unless $\alpha=0$.

\subsection{Quantum Effect}\label{Sec:degenerate_vacua}

In the previous subsections, we have solved the unitarity paradox by treating $\alpha$ classically. 
Here we consider how the above picture is modified by the quantum effect.

If there are no couplings between $\alpha$ and the massless fermions, from quantum mechanics on $\mathbb{S}^1$,
we know that the ground state is the superposition of different value of $\alpha$ with equal weight.
However, given the boundary-bulk coupling, the situation is not clear,\footnote{For finite $L$, the states with different $\alpha$ may not be degenerate in energy due to the Casimir energy.
In this case, we believe that $\alpha$ is localised.
In the limit $L\to\infty$, the states with different $\alpha$ are degenerate in energy, where the situation is not clear.} and there are two possibilities.
\begin{itemize}
    \item When the boundary-bulk coupling is large, it would be possible that the ground-state wavefunction of $\alpha$ has a localised profile, \ie, $\langle \mathrm{e}^{\i \alpha}\rangle\neq 0$. In the context of condensed matter physics, this is called (de)localisation of mobility \cite{Caldeira:1982iu,Caldeira:1982uj,Polychronakos:1992zn,Callan:1989mm}. 
    For example, 2d boundary conformal field theory considered in Ref.~\cite{Callan:1994ub} exhibits this behaviour.
    There is a quantum mechanical degree of freedom living on the boundary, and the boundary potential has several minima with the same potential energy.
    Thus,  there are the degenerate vacua classically, while the vacuum is unique quantum mechanically  if there are no couplings to the bulk.
    However, the authors of Ref.~\cite{Callan:1994ub} showed that, in the presence of couplings to the bulk, 
    the vacua become distinguishable at the quantum level and
    the degenerate vacua appear
    both classically and quantum mechanically,
    by explicitly computing the S-matrix element. 
    If our monopole-fermion system is also localised, then the solution in the previous subsection quantum mechanically applies as it is.
    Reference~\cite{Yegulalp:1994eq} considers a system close to our system, and shows that the system exhibits the localisation when $E\ll 1/I \ll 1/r_0$.
    Reference~\cite{Yegulalp:1994eq} also reproduces the dyon boundary condition considered in this paper, where the author computes the boundary entropy correctly.
    This is also consistent with the computation of the boundary entropy in Ref.~\cite{Smith:2020rru}. When the dyon boundary condition is imposed at both ends, there are $g_R^2=N/2$ sectors at the boundary. This number coincides with the number of the possible change of $\alpha$ $\alpha_\text{out}-\alpha_\text{in}=0,\frac{4\pi}{N},\frac{8\pi}{N},\cdots, \frac{4\pi}{N}(\frac{N}{2}-1)$.
    \item On the other hand, it is also possible that the wavefunction of $\alpha$ is not localised, since we are not aware of the study of the localisation exactly same as our system. Then, we should consider the superposition of $\alpha$ from different magnetic monopoles.
    Nevertheless, since the solution is valid for fixed $\alpha$, this must be valid for the superposition of $\alpha$. 
    {In this case, as an initial fermion, we can not consider the boundary condition \eqref{Eq:bc_chi_tilde} with fixed $\alpha$. Rather, we should consider the superposition of the boundary condition.
    Since the fermion number is related to $\alpha$ as in Eq.~\eqref{Eq:fermionnumber}, the superposition of $\alpha$ indicates that  we can not consider the initial fermion with the definite number. 
    The initial fermion is the superposition of the integer and fractional (as well as irrational) fermion numbers.
    In this picture, monopole-fermion scattering for $N=4$ would be understood in the following way. The initial fermion (on top of other boundary conditions) contains both periodic and anti-periodic boundary conditions.
    The periodic and anti-periodic components are scattered into anti-periodic and periodic components, respectively. In this way, the final state is again the superposition of $\alpha$, and the unitarity paradox would be solved.
    }

\end{itemize}
It is interesting to explore which is correct by studying the system analytically/numerically, which we leave for future study.
We emphasise here that the puzzle itself is solved independent of the quantum effect.

\section{Summary and Discussions}\label{sec:summary}

In this paper, we have proposed a solution to the unitarity paradox of the monopole-fermion scattering.
A crucial observation is that the Fock space of the fermions depends on the rotor degree of freedom of the monopole and changes by the scattering.
The problem is solved by realising that the fermions with the twisted boundary condition possess the fractional fermion number.
This provides a 4d interpretation of the 2d semiton state.

{A novelty of our proposal is the monopole internal state parameterised by the rotor degree of freedom. 
It is known that the rotor is related to the large gauge transformation of $U(1)_{\rm gauge}$~\cite{Harvey:1996ur}.
Since the degenerate vacua and memory effect in QED associated with the large gauge transformation are found recently~\cite{Strominger:2017zoo},
it is curious if our proposal is connected to that story. 
In particular, it is interesting if we could find the memory effect associated to the monopole-fermion scattering (see Refs.~\cite{Zeldovich:1974gvh,Braginsky:1985vlg,1987Braginskii_Thorne,Strominger:2014pwa} for the gravitational memory effect, and Refs.~\cite{Bieri:2013hqa,Susskind:2015hpa,Hamada:2017bgi,Hamada:2018cjj} for the electromagnetic memory effect).
}

The remaining of this paper is devoted to discussions.
In Sec.~\ref{sec:odd_N}, we comment on the case where the flavour number $N$ is odd.
The monopole in the context of $SU(5)$ GUT is discussed in Sec.~\ref{sec:GUT}.
The relations with other proposed solutions to the unitarity paradox are discussed in Sec.~\ref{sec:comments}.

\subsection{Odd \texorpdfstring{$N$}{N} cases}\label{sec:odd_N}

As we have argued, in the context with the $SU(2)$ gauge theory coupled to $N$-doublet Weyl fermions in the fundamental representation, $N$ must be even due to the Witten anomaly \cite{Witten:1982fp}.

However, the situation with $N=$ odd can be realised by the following setup. Let us consider an $SU(2)$ gauge theory coupled to $N$-triplet Weyl fermions in the adjoint representation as a UV complete system,
and assume that a Dirac monopole is produced in the IR theory with an adjoint Higgs.
Since the Witten anomaly is absent, $N$ can be not only even but also odd.
However, the $U(1)$ electric charge of the fermions is twice the minimum charge allowed by the Dirac-Zwanziger quantisation condition.
We leave the further investigation of the monopole-fermion scattering in this setup for future work.
See Refs.~\cite{McGreevy:2011if,Sato:2022vii} for related discussions.

\subsection{Monopoles in four-dimensional GUTs}\label{sec:GUT}
The monopole in the context of the $SU(5)$ GUT corresponds to the $N=4$ case, where the generator of $SU(2)$, $\bm{T}$, 
is embedded into the generator of $SU(5)$ as \cite{Dokos:1979vu,Schellekens:1983yc}
\begin{align}
    \bm{T}=\begin{pNiceMatrix}
          0&0&0&0&0\\
          0&0&0&0&0\\
          0&0&\Block{2-2}<>{\displaystyle \frac{\bm{\sigma}}{2}}&&0\\
          0&0&&&0\\
          0&0&0&0&0\\
    \end{pNiceMatrix}\,.
\end{align}
Here the first three and last two components correspond to $SU(3)_C$ and $SU(2)_L$ gauge groups of the Standard Model, and $\bm{\sigma}$ is the Pauli matrix.
After developing the vacuum expectation value of the adjoint Higgs, the unbroken $U(1)$ is
\begin{align}
    2\,T^3 = -\frac{1}{\sqrt{3}}\lambda^8-\frac{1}{2}\tau^3-\frac{1}{2}Y \,,
\end{align}
where $\lambda^{8}$ is one the Cartan subalgebra of $SU(3)_C$, $\tau^3$ is the subalgebra of $SU(2)_L$, and $Y$ is the generator of $U(1)_Y$ in the Standard Model. Explicitly, these generators are expressed as
\begin{align}
\begin{aligned}
\lambda^8 &= \mathrm{diag}\left(\frac{1}{\sqrt{3}},\frac{1}{\sqrt{3}},-\frac{2}{\sqrt{3}},0,0 \right)\,, \\
\tau^3 & = \mathrm{diag}(0,0,0,1,-1)\,, \\
Y&= \mathrm{diag}\left(-\frac{2}{3},-\frac{2}{3},-\frac{2}{3},1,1\right) \,.
\end{aligned}
\end{align}
Since the QED charge $Q_\text{QED}$ is given by $Q_\text{QED}=(\tau^3+Y)/2$, the unbroken $U(1)$ charge becomes
\begin{align}
    2\,T^3=-Q_{\lambda^8}-Q_\text{QED} \,,
\end{align}
where 
\begin{align}
    Q_{\lambda^8}:= \frac{\lambda^8}{\sqrt{3}} 
    =
    \begin{dcases}
    \frac{1}{3} & \text{for the colour index $1,2$} \,, \\
    -\frac{2}{3} & \text{for the colour index $3$} \,.
    \end{dcases}
\label{eq:Q8}\end{align}
In the left-handed basis, the Weyl fermion $SU(2)$ doublets ($=$ $U(1)$ Dirac fermions) are
\begin{align}
\begin{pmatrix}
\psi_{a}^{+} \\
\psi_{a}^{\prime -}
\end{pmatrix}=
\begin{pmatrix}
e_\text{L}\\
\left(d_\text{R}\right)^\dagger_3
\end{pmatrix}_{a=1},\quad
\begin{pmatrix}
\left(d_\text{L}\right)_3\\
e_\text{R}^\dagger
\end{pmatrix}_{a=2},\quad
\begin{pmatrix}
\left(u_\text{R}\right)^\dagger_1\\
\left(u_\text{L}\right)_2
\end{pmatrix}_{a=3},\quad
\begin{pmatrix}
\left(u_\text{R}\right)^\dagger_2\\
\left(u_\text{L}\right)_1
\end{pmatrix}_{a=4},
\label{eq:doublet}
\end{align} 
see, \eg, Ref.~\cite{Callan:1982ac}, where the upper and lower components correspond to $2T^3=\pm1$, and the colour index is $1,2,3$.
Note that the $U(1)$ charges of the other fermions ($\nu_{\text{L}}$, $(u_{\text{L}})_3$, $(u_{\text{R}})^\dag_3$, $(d_{\text{L}})_{1,2}$, and $(d_{\text{R}})^\dag_{1,2}$) are zero, so they do not interact with the monopole.

Here, we argue that, in addition to the $2T^3$ charge, all the charges (Gell-Mann matrices $\lambda^3$ and $\lambda^8$ corresponding to Cartan components, and $Q_\text{QED}$) are preserved by the scattering, as it should be.
The $\lambda^3$ charge is 
\begin{align}
Q_{\lambda^3} = 
\begin{dcases}
     \frac{1}{2}\,, & (c=1) \,, \\ 
     -\frac{1}{2}\,, & (c=2) \,, \\ 
     0\,, & (c=3) \,,
\end{dcases}
\label{eq:lambda3}
\end{align}
while the $\lambda^8$ charge is given in Eq.~\eqref{eq:Q8}.
The $Q_{\lambda^3}$ charge is a part of the global $SU(4)$ symmetry, and is shown to be conserved in Ref.~\cite{Maldacena:1995pq}.
By looking Eq.~\eqref{eq:doublet}, the $Q_{\lambda^8}$ charge before and after the scattering of the massless fermions off a $\mm=1$ monopole is given by
\begin{align}
    (Q_{\lambda^8})_\text{in}=\frac{2}{3}n_2^\text{in}+\frac{1}{3}n_3^\text{in}+\frac{1}{3}n_4^\text{in} \,, \qquad 
    (Q_{\lambda^8})_\text{out}=-\frac{2}{3}n_1^\text{out}-\frac{1}{3}n_3^\text{out}-\frac{1}{3}n_4^\text{out} \,,
\end{align}
where the incoming (outgoing) states correspond to the upper (lower) components of Eq.~\eqref{eq:doublet}, and $n_a$ is the fermion number of $a$-th flavour, as defined in Eq.~\eqref{eq:n_def}.
By substituting Eq.~\eqref{eq:change_fermion_number} into the above equation, we can see  $(Q_{\lambda^8})_\text{in}=(Q_{\lambda^8})_\text{out}$.
A similar calculation, as expected, also shows $(Q_{\text{QED}})_\text{in}=(Q_{\text{QED}})_\text{out}$ via Eq.~\eqref{eq:change_fermion_number}.
The conservation of the all Cartan charges implies that the semiton description is correct even in the context of monopole scattering in the $SU(5)$ GUT (with vanishing Yukawa couplings and single generation).

Summary of the $SU(5)$ GUT monopole scattering:
A process of the monopole catalysis of proton decay is described by
\begin{align}
  M^{-} + (u_{\text{L}})_1 + (u_{\text{L}})_2 \to M^{-} + (e_{\text{L}})^\dag + (d_{\text{L}})_3^\dag\,,
  \label{Eq:protondecay}
\end{align}
where $M^{-}$ is a $\mm=-1$ monopole.
The total fermion number changes by $\Delta F =-4$.
On the other hand, the monopole-electron scattering is described as 
\begin{align}
  M^{+} + (e_{\text{L}}) \to M^{+} +\frac{e_{\text{R}}}{2} + \frac{(u_{\text{L}})_1^\dag}{2} + \frac{(u_{\text{L}})_2^\dag}{2} + \frac{(d_{\text{R}})_3^{\dag}}{2}\,,
   \label{Eq:monopoleelectron}
\end{align}
where $M^{+}$ is a $\mm=1$ monopole
and $\psi_a/2 ~(\bar\psi_{\bar a}/2)$ denotes the fermion state whose fermion number  as well as gauge charges is half of $\psi_a \,(\bar\psi_{\bar a})$.
The fractional fermion number 
in the final state is a result of the vacuum transition through Eq.~\eqref{Eq:alpha_change},
and the total fermion number changes by $\Delta F =-2$.
Notice that, in both processes, all gauge charges as well as the unbroken $U(1)$ charge ($2 T^3$) are conserved.

\subsection{Comments on other proposals}\label{sec:comments}
To the best of our knowledge, there are three recent proposals \cite{Kitano:2021pwt,Csaki:2021ozp,Brennan:2021ewu} as a solution to the unitarity paradox on 4d.
Here, we briefly comment on these proposals and their possible relation with our proposal.
Although some proposals have similarities with our proposal, the precise relationship is not clear. 
It is interesting to investigate the relation in detail.
\begin{itemize}
\item Kitano-Matsudo \cite{Kitano:2021pwt}

They argue for the existence of a new fermion whose charge is the same as $X$ in Eq.~\eqref{???}.
The new fermion, as well as the original fermions, are described by the 4d solitons called \emph{pancakes}.
They claim that in the presence of the monopole, the spectra of the massless QED are doubled.
This is in contrast with out solution where the doupling of the spectra does not occur.

\item Brennan \cite{Brennan:2021ewu}

Based on the results in Ref.~\cite{Brennan:2021ucy}, the effective 2d theory is derived by integrating out massive modes from 4d Lagrangian. 
The semiton is interpreted as the soft radiation of the massless charged fermions.
If we interpret this soft radiation as fermion zero modes, this is similar to our solution. 
However, in our case, the fermion zero energy level appears only after the scattering.

\item Cs\'aki-Shirman-Telem-Terning \cite{Csaki:2021ozp}

The statement is completely different from this paper and even from others.
Using a spinor-helicity formalism developed in Refs.~\cite{Csaki:2020inw,Csaki:2020yei}, they conclude that the multi-particle states cannot be written as tensor products of the one-particle states. 
More specifically, each angular momentum of the multi-fermions after the monopole scattering can be {\em entangled} that can be expressed by {\em pairwise helicity}.
The classification of the incoming and outgoing fermions based on the one-particle state in Table~\ref{tab:charge} would not be appropriate. 
Based on the new formalism, they show that the monopole-electron scattering is (cf. Eq.~\eqref{Eq:monopoleelectron})
\begin{align}
  M^{+} + e_{\text{L}} \to M^{+} + (u_{\text{L}})_1^\dag + (u_{\text{L}})_2^\dag + (d_{\text{L}})_3^\dag\,,
\end{align}
where all gauge charges are conserved again.
These final states are different from our final state.
From kinematics, both final states can appear in principle. It is interesting to explore which final state dominantly appears at the low energy scattering we are interested in.

\end{itemize}

\acknowledgments

The authors thank Yuji Tachikawa for collaboration in the early stages of this work and valuable discussions and thank Masataka Watanabe for important input and fruitful discussions during this collaboration. 
We also thank Ryuichiro Kitano, Yui Hayashi, Ryutaro Matsudo, Shigeki Sugimoto, Juven Wang, Satoshi Yamaguchi and Kazuya Yonekura for useful discussions.
The work of Y.H. was supported by Japan Society for the Promotion of Science (JSPS) Overseas Research Fellowships.
At the final stages of the work, Y.H. was supported by MEXT Leading Initiative for Excellent Young Researchers Grant Number JPMXS0320210099.
The work of T.K. is supported by the JSPS Grant-in-Aid for Early-Career Scientists (No.\,19K14706)
and by the JSPS Core-to-Core  Program (Grant No.\,JPJSCCA20200002). 
The work of Y.S. was supported by JSPS KAKENHI Grant (No.\,20K22344).
Y.S. is the Yukawa Research Fellow supported by Yukawa Memorial Foundation.

\bibliographystyle{JHEP}
\bibliography{reference}

\end{document}